\documentclass[journal]{IEEEtran}  % Comment this line out

\IEEEoverridecommandlockouts                             
\usepackage[mode=buildmissing]{standalone} % mode=image|tex, mode=build, buildmissing

\usepackage{amsmath} 
\usepackage{amssymb}
\usepackage{amsthm}
\usepackage{bm}
\usepackage{color}
\usepackage{verbatim}
\usepackage{booktabs}
\usepackage{amsmath}
\usepackage{balance}
\usepackage{graphicx}
\usepackage{tabu}
\usepackage[font={footnotesize}]{caption}
\usepackage{array, ltablex, multirow,ragged2e}
\usepackage{hyperref}
\usepackage{longtable}
\usepackage{mdframed}
\usepackage{subfig} 
\usepackage{dsfont}
\usepackage{cite}
\usepackage{float}
\usepackage{algorithm}
\usepackage[noend]{algpseudocode}
\usepackage{tikz,pgfplots}
\usepgfplotslibrary{fillbetween}
\usepgfplotslibrary{groupplots}
\makeatletter
\def\BState{\State\hskip-\ALG@thistlm}
\makeatother

\theoremstyle{plain}

\theoremstyle{remark}

\theoremstyle{plain}

\graphicspath{{../figures/}}

\pgfplotsset{compat=newest}%
\usetikzlibrary{positioning,matrix,shapes.multipart,shapes.misc,spy}%
\tikzstyle{annotation}=[fill=white]%
\newcommand{\mat}[1]{\boldsymbol{#1}}
\newcommand{\vect}[1]{\boldsymbol{#1}}

\newcommand{\transpose}{\mathsf{T}}
\newcommand{\Ctranspose}{\mathsf{H}}
\newcommand{\define}{\triangleq}

\DeclareMathOperator*{\argmax}{arg\,max}
\DeclareMathOperator*{\argmin}{arg\,min}

\usetikzlibrary{shapes.arrows}
\usetikzlibrary{spy}

\tikzset{%
	partial ellipse/.style args={#1:#2:#3}{%
		insert path={+ (#1:#3) arc (#1:#2:#3)}%
	}%
}%

\tikzstyle{style A}=[black, mark=diamond*, mark options={solid, fill=white, mark size=2.0pt}, solid]%
\tikzstyle{style B}=[color=blue, mark=*, mark options={solid, fill=white, mark size=1.5pt}, dotted]%
\tikzstyle{annotation}=[fill=white]%

\tikzstyle{star marker}=[mark=star, mark options={solid, scale=1}]
\tikzstyle{diamond marker}=[mark=diamond*, mark options={solid, fill=white, mark size=2.0pt}]
\tikzstyle{triangle marker}=[mark=triangle*, mark options={solid, scale=0.8}]
\tikzstyle{square marker}=[mark=square*, mark options={solid, scale=0.8}]
\tikzstyle{circle marker}=[mark=*, mark options={solid, scale=1}]
\tikzstyle{autoencoder}=[thick, color=red, triangle marker, dashed]%
\tikzstyle{baseline}=[thick, color=blue, square marker, solid]%
\tikzstyle{ibaseline}=[thick, color=green!50!black, no markers, solid]%

\newcommand{\rr}[2]{%
    \begin{tabular}{@{}c@{}}#1 \\ #2\end{tabular}
}%

\newcommand{\HW}[1]{{\color{red}{[HW: #1]}}}

\newcommand{\RevA}[1]{%
    {%\color{red}%
	#1%
	}%
}%

\newcommand{\RevB}[1]{%
    {%\color{blue}%
	#1%
	}%
}%

\title{Benchmarking and Interpreting End-to-end Learning of MIMO and Multi-User Communication}

\author{%
Jinxiang Song, \emph{Student Member, IEEE}, 
Christian H\"{a}ger, \emph{Member, IEEE}, 
Jochen Schr\"{o}der, \emph{Member, IEEE}, 
Timothy J.~O'Shea, \emph{Senior Member, IEEE}, 
Erik Agrell, \emph{Fellow, IEEE}, 
Henk Wymeersch, \emph{Senior Member, IEEE}
\thanks{Parts of this paper have been presented at the \emph{IEEE Global Communications Conference (GLOBECOM)}, Taipei, Taiwan, 2020.}%
\thanks{This work was supported by the Knut and Alice Wallenberg Foundation, grant No.~2018.0090, and the Swedish Research Council under grant  No.~2018-0370. 
The work of Christian~H\"ager was supported by the European Union's Horizon 2020 research and innovation programme under the Marie Sk\l{}odowska-Curie grant No.~749798. \emph{(Corresponding author: Jinxiang Song)}}
\thanks{%
Jinxiang Song, Christian H\"{a}ger, Erik Agrell, and Henk Wymeersch are with the Department of Electrical Engineering, Chalmers University of Technology, 41296 Gothenburg, Sweden (emails: \{jinxiang, christian.haeger, agrell, henkw\}@chalmers.se).}
\thanks{%  
Jochen Schr\"{o}der is with the Department of Microtechnology and Nanoscience, Chalmers University of Technology, 41296 Gothenburg, Sweden (email: jochen.schroeder@chalmers.se)}
\thanks{%
Timothy J.~O'Shea is with the Bradley Department of Electrical and Computer Engineering, Virginia Tech and DeepSig, Inc., Arlington, VA 22203, USA (email: oshea@vt.edu). 
}%
}%

\begin{document}

\maketitle

%%%%%%%%%%%%%%%%%%%%%%%%%%%%%%%%%%%%%%%%%%%%%%%%%%%%%%%%%%%%%%%%%%%%%%%%%%%%%%%%
\begin{abstract}
End-to-end autoencoder (AE) learning has the potential of exceeding the performance of human-engineered transceivers and encoding schemes, without a priori knowledge of communication-theoretic principles. 
In this work, we aim to understand to what extent and for which scenarios this claim holds true when comparing with fair benchmarks. 
Our particular focus is on \RevA{memoryless} multiple-input multiple-output (MIMO) and multi-user (MU) systems. Four case studies are considered: two point-to-point (closed-loop and open-loop MIMO) and two MU scenarios (MIMO broadcast and interference channels).
%For the point-to-point scenarios, it is shown that previously observed performance gains of AE-based communication can be ascribed to an implicitly learned geometric shaping and bit and power allocation, not to learning new spatial encoders. 
For the point-to-point scenarios, we explain some of the performance gains observed in prior work through the selection of improved baseline schemes that include geometric shaping as well as bit and power allocation. 
For the MIMO broadcast channel, we demonstrate the feasibility of a novel AE method with centralized learning and decentralized execution. 
Interestingly, the learned scheme performs close to nonlinear vector-perturbation precoding and significantly outperforms conventional zero-forcing. 
Lastly, we highlight potential pitfalls when interpreting learned communication schemes. 
In particular, we show that the AE for the considered interference channel learns to avoid interference, albeit in a rotated reference frame. 
After de-rotating the learned signal constellation of each user, the resulting scheme corresponds to conventional time sharing with geometric shaping.
%For each scenario, we provide explicit descriptions as well as open-source implementations of the selected neural-network architectures and benchmark algorithms. 
\end{abstract}


%%%%%%%%%%%%%%%%%%%%%%%%%%%%%%%%%%%%%%%%%%%%%%%%%%%%%%%%%%%%%%%%%%%%%%%%%%%%%%%%
\section{Introduction}

Demand for higher data rates has led to the continued development of ever more performant wireless communication systems.
One of the most important developments has been multiple-input multiple-output (MIMO) transmission \cite{Paulraj2004}, where information across multiple antenna elements is encoded using spatial-multiplexing or spatial-diversity schemes to enhance throughput and reliability of  communication systems. 
Conventional MIMO communication systems are often classified as closed-loop or open-loop. 
In open-loop systems, channel state information (CSI) is only available at the receiver, while in closed-loop systems, the transmitter also has access to CSI (either through explicit feedback or via channel reciprocity). 
Several approaches have been used for both open-loop and closed-loop systems, including maximum-likelihood detection, zero-forcing (ZF) precoding, minimum mean-square-error (MMSE) equalization, space-time block coding, and singular value decomposition (SVD) with waterfilling \cite[Chapter 11]{WilWilBig16}. 

Recent years have witnessed a resurgence of interest in machine-learning (ML) techniques for communication systems.
Most work has focused on supervised learning for \emph{specific functional blocks} such as modulation recognition \cite{OShea2016}, MIMO detection \cite{Samuel2017, Samuel2019, Jeon2017, Nguyen2018}, MIMO channel estimation \cite{He2018}, and channel decoding \cite{Gruber2017, Nachmani2018}.
These ML-based methods have led to algorithms that often perform better or exhibit lower complexity than model-based algorithms. 
In contrast to focusing on specific functional blocks, \emph{end-to-end learning} has been proposed to optimize the transmitter and receiver jointly \cite{OShea2017}. 
The workhorse of end-to-end learning is the autoencoder (AE), which employs two neural networks (NNs) to encode and decode messages into a learned latent representation which passes through a physical communication channel.
This method has been successfully applied to a wide variety of channels, including, e.g., linear wireless \cite{Doerner2018, He2019}, and nonlinear optical \cite{Shen2018ecoc, Karanov2018} ones. 
In cases where no differentiable channel model is available, a surrogate channel can first be learned \cite{Ye2018, OShea2019} or the transmitter can be designed as a reinforcement-learning agent \cite{Aoudia2019}, which can be trained even with limited reward feedback \cite{Song2020}. 

\newcolumntype{C}[1]{ >{\centering\arraybackslash} m{#1} }
\newcommand{\alert}[1]{\textbf{#1}}
\setlength\tabcolsep{3.5pt} % default value: 6pt
\renewcommand{\arraystretch}{1.25}

\begin{table*}[t]
\centering
\caption{Overview of the considered scenarios, best-performing baseline schemes, and high-level conclusions in this and previous works. (AE: autoencoder, STBC: space-time block code, SVD: singular value decomposition, GS: geometric shaping)}
\begin{tabular}{c | C{3.8cm} C{3.8cm} C{3.8cm} C{3.8cm}}
    \toprule
    & (i) open-loop MIMO & (ii) closed-loop MIMO & (iii) MIMO broadcast & (iv) interference channel \\
    \midrule
    reference(s) &  \cite{OShea2017Physical, OShea2017Deep} &  \cite{OShea2017Physical, OShea2017Deep} & \cite{Pathapati2020}$^{*}$ & \cite{OShea2017} \\
    baseline & Alamouti STBC w/ QAM & SVD w/ QPSK and equal power & Tomlinson--Harashima & time sharing w/ QAM \\
    conclusion & \emph{AE outperforms baseline} & \emph{AE outperforms baseline} & \emph{AE outperforms baseline} & \emph{AE outperforms baseline} \\
    \midrule
    our baseline & Alamouti STBC w/ GS & SVD w/ GS, bit$\,$\&$\,$power loading & vector-perturbation precoding & time sharing w/ GS \\
    our conclusion & \emph{AE matches baseline} & \emph{baseline outperforms AE}${}^\dagger$ & \emph{baseline outperforms AE} & \emph{baseline outperforms AE}\\
    \bottomrule
\end{tabular}
\RaggedRight

\smallskip

$\quad^{*}$Independently proposed in this paper (preliminary results presented in \cite{Song2020Globecom}).

$\quad^\dagger$AE outperforms baseline for certain channel singular values, see Sec.~\ref{sec:closed-loop_results}.
    
    \label{tab:overview}
\end{table*}

In this paper, we consider the application of end-to-end learning to MIMO systems assuming both point-to-point and multi-user (MU) transmission scenarios. 
For these applications, there has been limited treatment of AEs. 
In \cite{OShea2017Physical}, open-loop and closed-loop MIMO were studied, leading to better performance than the selected benchmark methods. 
In the extension \cite{OShea2017Deep}, finite quantization of the CSI was considered, which was demonstrated to further improve performance under some conditions.
While \cite{OShea2017Physical, OShea2017Deep} have shown promising performance of AE-based MIMO communication, the proposed systems were trained under some nonstandard assumptions, specifically regarding CSI availability at the receiver and power normalization at the transmitter, as explained in more detail below.
Besides \cite{OShea2017Physical, OShea2017Deep}, MIMO AEs were also studied in \cite{Wang2020, ElMossallamy2019} for the noncoherent case, where neither the transmitter nor the receiver has access to CSI. 
Regarding MU communication, the authors in \cite{OShea2017} have shown that the AE framework can be extended to include multiple transmitter--receiver pairs. 
They considered a conventional Gaussian interference channel and showed that the learned communication scheme achieves better performance than the selected time-sharing baseline. 

In this paper, we build on the approaches proposed in \cite{OShea2017Physical, OShea2017Deep, OShea2017} with the aim to better understand what performance gains can be achieved by AE-based MIMO and MU systems under more realistic training assumptions when compared to fair benchmarks. \RevA{To that end, the channel models considered are assumed to be memoryless, as in \cite{OShea2017Physical, OShea2017Deep, OShea2017}.}
Moreover, we also provide additional interpretations of the learned communication schemes.
A particular emphasis in this work is placed on selecting baseline schemes with geometric shaping (GS), see, e.g., \cite{Foschini1973}. 
Shaped modulation formats for Gaussian channels are also readily available in open databases \cite{codes.se}. 
Our main contributions in this work are as follows:
\begin{itemize}
    \item For the MIMO systems in \cite{OShea2017Physical, OShea2017Deep}, we analyze and evaluate the corresponding AEs under more standard training assumptions. 
    In particular, while CSI in \cite{OShea2017Physical,OShea2017Deep} was assumed to be estimated at the receiver, it was not actually used as a receiver input. 
    Moreover, power normalization was applied after the channel-matrix multiplication (cf.~\cite[Eqs.~(2), (3)]{OShea2017Deep}), which cannot be done in practical systems. 
    By contrast, our AEs always use the CSI as an additional receiver input and power normalization is performed prior to the channel. 
    Additionally, reproducible open-source implementations of our AEs and benchmark schemes are also provided.\footnote{The complete source code to reproduce all results in this paper is available at \url{https://github.com/JSChalmers/DeepLearning_MIMO.git}}
    
    \item We then explain some of the performance gains obtained by the trained AEs through the selection of improved baseline schemes compared to \cite{OShea2017Physical,OShea2017Deep}. 
    In particular, for open-loop MIMO, we show that previously observed performance gains can be partially attributed to an implicit GS of the underlying signal constellation. 
    For closed-loop MIMO, we use an SVD-based benchmark similar to \cite{OShea2017Physical,OShea2017Deep}, but augment it through GS as well as additional bit and power allocation. 
    This closes the performance gap to the AE, indicating that the ML-based solution learns to implement similar functionalities in a data-driven fashion.
    
    \item We then propose and analyze a novel AE system for a MIMO broadcast scenario, where a single multi-antenna transmitter sends information to multiple single-antenna users.\footnote{This scenario was suggested as a possible extension in \cite[Sec.~V]{OShea2017Deep}.} 
    For such a system, we extend the training methodology in \cite{OShea2017Deep} to account for the joint loss function of all users. 
    The resulting AE is shown to provide performance between nonlinear vector-perturbation precoding \cite{Hochwald2005} and conventional transmitter ZF, significantly outperforming the latter over a wide range of signal-to-noise ratios (SNRs). 
    In parallel to our work, a similar scenario was also recently considered in \cite{Pathapati2020}. 
    This work is discussed in more detail in Sec.~\ref{sec:mu-mimo_results}. 
    
    \item Lastly, we revisit the interference-channel scenario in \cite{OShea2017} where significant performance gains were demonstrated by AE-based communication compared to the considered time-sharing baseline scheme. 
    After augmenting the time-sharing scheme with a geometrically-shaped signal constellation, we find that the improved baseline performs similarly to (and in some cases even better than) the AE. 
    The improved baseline also allows us to provide an additional theoretical interpretation of the learned AE scheme in terms of a ``rotated'' time-sharing scheme. 

\end{itemize}
An overview of the considered scenarios including the best-performing baseline schemes and high-level conclusions can be found in Table~\ref{tab:overview}.
We note that the underlying assumptions for each scenario (e.g., the fading model or the number of transmit/receive antennas) are consistent with prior work, which allows us to make direct comparisons to the corresponding results. 
We comment on some of the limitations of these assumptions in Sec.~\ref{sec:conclusion}. 

The remainder of the paper is structured as follows. 
In Sec.~\ref{sec:siso_ae}, a brief introduction to AE-based communication is given. 
The four scenarios listed in Table~\ref{tab:overview} are then studied in Secs.~\ref{sec:open-loop} (open-loop MIMO), \ref{sec:closed-loop} (closed-loop MIMO), \ref{sec:mu-mimo} (MIMO broadcast), and \ref{sec:ic} (interference channel), where each section contains a detailed description of the baseline scheme(s), AE implementation, as well as numerical results and a discussion. 
Finally, the paper is concluded in Sec.~\ref{sec:conclusion}.

% information-theoretic literature: 
% - point-to-point MIMO \cite{Telatar1999}
% - MIMO broadcast \cite{Caire2003, Vishwanath2003, Weingarten2006}
% - interference channel \cite{Han1981, Zamir2002, Etkin2008, Cadambe2008}
% Etkin2008 consider 2-user IC

\emph{Notation:} $\mathbb{Z}$, $\mathbb{R}$, and $\mathbb{C}$ denote the sets of integers, real numbers, and complex numbers, respectively.
We use boldface letters to denote vectors and matrices (e.g., $\vect{x}$ and $\mat{A}$).
$(\cdot)^\transpose$ and $(\cdot)^\Ctranspose$ denote transpose and conjugate transpose, respectively. 
For a vector $\vect{x} = [x_1, \ldots, x_n]^\transpose$, $[\vect{x}]_i = x_i$ returns the \mbox{$i$-th} element of $\vect{x}$, $\|\vect{x}\|^2 = \sum_{i=1}^n |x_i|^2$ denotes the squared Euclidean norm, and $\mathrm{diag}(\vect{x})$ is the matrix whose diagonal entries are the elements of $\vect{x}$.  
A matrix $\mat{X}$ is converted to a vector by stacking the columns, which is denoted by $\mathrm{vec}(\mat{X})$. 
$\mat{I}_n$ is the $n \times n$ identity matrix.
$[a,b]^M$ is the $M$-fold Cartesian product of the interval $[a,b]$. 
$\mathcal{CN}(\boldsymbol{x};\boldsymbol{\mu},\boldsymbol{\Sigma})$ denotes the distribution of a proper complex Gaussian random vector with mean $\boldsymbol{\mu}$ and covariance matrix $\boldsymbol{\Sigma}$, evaluated at $\boldsymbol{x}$ ($\boldsymbol{x}$ may be omitted to represent the entire distribution). 
Lastly, $\mathbb{E}\{\cdot\}$ denotes expected value. 

% Element-wise vector multiplication is denoted by $\circ$, i.e., $\vect{x} \circ \vect{y} = \diag(\vect{x}) \vect{y}$. 

\section{Autoencoder-based Communication Systems}
\label{sec:siso_ae}

In this section, we start by briefly reviewing AE-based communication assuming transmission over the memoryless (complex-valued) additive white Gaussian noise (AWGN) channel 
\begin{align}
    \label{eq:awgn}
    \vect{y}_{k} = \vect{x}_{k} + \vect{n}_{k},
\end{align}
where $\vect{x}_k, \vect{y}_k \in \mathbb{C}^{N_B}$ are the channel input and output vector in the $k$-th transmission block, $N_B$ denotes the number of channel uses per block, and $\boldsymbol{n}_{k} \sim \mathcal{CN}(\boldsymbol{0},N_0\boldsymbol{I}_{N_B})$ is independent and identically distributed (i.i.d.)~Gaussian noise. 
The specific AE implementations for the considered MIMO and MU scenarios are then described in detail in the following sections. 

\subsection{Transmitter and Receiver Design}

AE-based end-to-end learning was originally proposed in \cite{OShea2017}. 
The general idea is to reinterpret the design of a communication system as a reconstruction task that jointly optimizes parameterized transmitters and receivers. 
To that end, the transceiver for the AWGN channel \eqref{eq:awgn} can be implemented by a pair of NNs $f_\tau: \mathcal{M}\to \mathbb{C}^{N_B}$ and $f_\rho:\mathbb{C}^{N_B}\to[0,1]^M$,  where $\mathcal{M} = \{1, 2, \ldots, M\}$ is the message set and $\tau$ and $\rho$ are the transmitter and receiver NN parameters, respectively.
More precisely, a message $m_k \in \mathcal{M}$ is first encoded as an $M$-dimensional ``one-hot'' vector, where the $m_k$-th element is $1$ and all the others are $0$. 
This vector is then used as the input to the transmitter NN.\footnote{\RevB{\textcolor{black}{In principle, other message encodings can also be used, see \cite{OShea2017} for details}, which are particularly important for large message sets.}} 
The NN is assumed to have $2 N_B$ output neurons, which form the real and imaginary part of the unnormalized transmit vector denoted by $\tilde{\vect{x}}_k = \tilde{f}_\tau(m_k) \in \mathbb{C}^{N_B}$.
The average transmit power is defined as $P_T = \mathbb{E}\{\|\vect{x}_k\|^2\} / N_B$ and enforced by a normalization layer such as
\begin{align}
    \label{eq:normalization}
    \vect{x}_k = \frac{\tilde{\vect{x}}_k \sqrt{N_B P_T}}{\sqrt{\frac{1}{M}\sum_{i=1}^M 
    \| 
    %\tilde{\vect{x}}_{k'} 
    \tilde{f}_\tau(i)
    \|^2}},
\end{align}
where $\vect{x}_k = f_\tau(m_k)$ denotes the entire transmitter mapping.
The vector $\vect{x}_k$ is then sent over the channel \eqref{eq:awgn} and the receiver NN processes the received vector $\vect{y}_k$ by generating an $M$-dimensional probability vector $\boldsymbol{q}_k=f_\rho(\vect{y}_k)$, where the components of $\boldsymbol{q}_k$ can be interpreted as the estimated posterior probabilities of the messages. 
Finally, the transmitted message is estimated according to $\hat{m}_k=\argmax_m[\boldsymbol{q}_k]_m$.

\subsection{End-to-end Training Procedure}

To optimize the transmitter and receiver parameters, it is important to have a suitable optimization criterion. Due to the fact that optimization relies on the empirical computation of gradients, a criterion like block error rate (BLER) $\mathrm{Pr}\{\hat{m}_k\neq m_k\}$ cannot be used directly.
Instead, a commonly used criterion is the categorical cross-entropy loss function \cite{OShea2017} defined by
\begin{equation}
\label{eq:ce_loss}
    %_{m_k,\vect{y}_k}
    \mathcal{J}_{\text{CE}}(\tau, \rho)=-\mathbb{E}\{ \log [f_{\rho}(\vect{y}_k)]_{m_k}\},
\end{equation}
where the dependence of $\mathcal{J}_{\text{CE}} (\tau,\rho)$ on $\tau$ is implicit through the distribution of the channel output $\vect{y}_k$, which is a function of the channel input $f_{\tau}(m_k)$. 
This loss function is also adopted for all scenarios in this paper, either directly or in the form of a weighted average (for cases involving multiple users), as explained in detail below.
\RevB{In practice, $\mathcal{J}_\text{CE}$ is usually approximated via Monte Carlo estimation. More specifically, a batch (or minibatch) of $B$ samples is randomly chosen in each gradient step and $\mathcal{J}_{\text{CE}}$ is approximated according to 
\begin{align}
    \hat{\mathcal{J}}_\text{CE} = - \frac{1}{B}\sum_{k=1}^{B}\log [f_{\rho}(\vect{y}_k)]_{m_k}.
\end{align}
}
Optimization of the NNs can be performed by minimizing $\hat{\mathcal{J}}_{\text{CE}}$ through the widely used Adam optimizer~\cite{Kingma2015}, or a variety of similar stochastic gradient descent optimizers.

\section{Open-Loop MIMO}
\label{sec:open-loop}

In this section, we consider an open-loop MIMO system where a transmitter with $N_T$ antennas sends sequences of messages to a transmitter with $N_R$ antennas. We note that for all scenarios in this paper, the information rate is always assumed to be fixed and forward error correcting coding is not considered.

\subsection{Background and Baseline Schemes}
\label{sec:open-loop:background}

The channel matrix at discrete time $k$ is denoted by $\boldsymbol{H}_k \in \mathbb{C}^{N_R \times N_T}$. 
The channel is drawn from a stationary distribution and is assumed to be block fading with duration $N_B \geq N_T$. 
In open-loop systems, CSI is available at the receiver but not at the transmitter. 
Conventional transmit approaches for open-loop MIMO systems include space-time block codes (STBCs) \cite{Alamouti1998, Tarokh1998, Tarokh1999}, which are described next. 

The transmitter generates $L$ messages, maps each message to a data symbol $s_{k,l} \in \Omega$ from a complex signal constellation $\Omega \subset \mathbb{C}$, and then encodes $\boldsymbol{s}_k=[s_{k,1},\ldots,s_{k,L}]^\transpose$ using a STBC with rate $L/N_B \le 1$. 
The resulting $N_B$ coded vectors of length $N_T$ are denoted by $\boldsymbol{X}_k=[\boldsymbol{x}_{k,1},\ldots,\boldsymbol{x}_{k,N_B}]$, with the property that $\mathbb{E}\{\boldsymbol{X}^\Ctranspose_k\boldsymbol{X}_k\}= P_T\boldsymbol{I}_{N_B}$, where $P_T$ is the total average transmit power, summed over all transmit antennas. 
If each of the $L$ complex data symbols corresponds to $\log_2(M)$ bits (i.e., one message), then the total bit rate is $r=L\log_2(M)/N_B$. 
The receiver observes
\begin{align}
    \boldsymbol{Y}_{k}=\boldsymbol{H}_k\boldsymbol{X}_{k} + \boldsymbol{N}_{k},% \quad p=1,\cdots,N_B,
    \label{eq:OLMIMO-obs}
\end{align}
where $\mathrm{vec}(\boldsymbol{N}_{k}) \sim \mathcal{CN}(\boldsymbol{0},N_0\boldsymbol{I}_{N_R N_B})$ is i.i.d.~Gaussian noise. 
The receiver then applies maximum-likelihood detection to $\boldsymbol{Y}_{k}=[\boldsymbol{y}_{k,1},\ldots, \boldsymbol{y}_{k,N_B}]$ according to
\begin{align}
\label{eq:open_loop_ml}
  \hat{\vect{s}}_k = \argmin_{\vect{s}_k \in \Omega^L} \| \mathrm{vec} (\mat{Y}_k - \mat{H}_k \mat{X}_k) \|^2,
\end{align}
which can be achieved through low-complexity linear processing \cite{Tarokh1999}. 
Other (less complex) receiver approaches for open-loop MIMO include ZF and MMSE detection, which are not considered here as they are suboptimal.

In this paper, we restrict ourselves to the Alamouti STBC \cite{Alamouti1998}, where $N_T=2$, $N_B=2$, $L=2$, with $r=\log_2(M)$. 
As an example, the Alamouti STBC for $N_R = 1$ is defined by the mapping
\begin{align}
   \vect{x}_{k,1} &= [s_{k,1}, s_{k,2}]^\transpose,\\
   \vect{x}_{k,2} &= [-s_{k,2}^*, s_{k,1}^*]^\transpose.
\end{align}
At the receiver, one may first form a combination of the two received symbols $\mat{Y}_k = [y_{k,1}, y_{k,2}]$ according to 
\begin{align}
    \tilde{s}_{k,1} &= h_{k,1}^* y_{k,1} + h_{k,2} y_{k,2}^* , \\
    \tilde{s}_{k,2} &= h_{k,2}^* y_{k,1} - h_{k,1} y_{k,2}^* ,
   % \tilde{\vect{s}}_k &= h_{k,1}^* y_{k,1} +  h_{k,2} y_{k,2}^*, \\
\end{align}
where $\mat{H}_k = [h_{k,1}, h_{k,2}]$ is the $1 \times 2$ channel matrix in this case. 
An optimal decision can then be made separately based on $\tilde{s}_{k,1}$ and $\tilde{s}_{k,2}$.

\subsection{Autoencoder Design and Training}

\label{open-loop-mimo}
\begin{figure}[t]
    \centering
    \includegraphics[width=8.7cm]{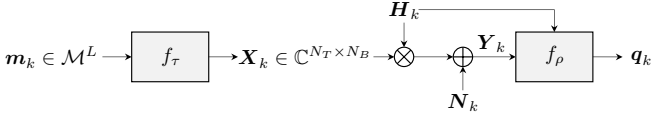}
    \caption{Open-loop MIMO channel AE, where the transmitter learns a rate $L/N_B$ code without CSI, while the receiver learns a decoder in the presence of CSI. The channel is drawn i.i.d.~from the underlying distribution.}
    \label{fig:open-loopSYSTEM}
\end{figure}

For an open-loop MIMO system with CSI available to the receiver, the AE implementation is visualized in Fig.~\ref{fig:open-loopSYSTEM}. 
The transmitter $f_\tau: \mathcal{M}^L\to \mathbb{C}^{N_T\times N_B}$ maps $L$ consecutive messages $\boldsymbol{m}_k = [m_1, \ldots, m_L]^\transpose \in \mathcal{M}^L$ to $N_B$ coded vectors according to
\begin{equation}
   \boldsymbol{X}_k = [\boldsymbol{x}_{k,1}, \ldots, \boldsymbol{x}_{k,
   N_B}]= f_\tau(\boldsymbol{m}_k),
\end{equation}
where $ \boldsymbol{x}_{k,p}$, $p=1, \ldots,N_B$, is a column vector of length $N_T$. An average power constraint according to $\sum_{p=1}^{N_B} \mathbb{E}\{\Vert \boldsymbol{x}_{k,p} \Vert ^2\} = N_B P_T$ is enforced through a normalization layer similar to \eqref{eq:normalization}. Inside $f_\tau(\cdot)$, an encoding of $\boldsymbol{m}_k$ to an $M^L$-dimensional one-hot vector is used. 
The receiver $f_{\rho}: \mathbb{C}^{N_R\times N_B} \times \mathbb{C}^{N_R\times N_T} \to [0,1]^{M^L}$ observes $\boldsymbol{Y}_{k}=[\boldsymbol{y}_{k,1},\cdots \boldsymbol{y}_{k,N_B}]$ as in \eqref{eq:OLMIMO-obs} and generates a probability vector $\boldsymbol{q}_k\in[0,1]^{M^L}$ according to 
\begin{align}
\label{eq:NN_decoder}
    \boldsymbol{q}_k=f_\rho(\boldsymbol{Y}_{k}, \boldsymbol{H}_k),
\end{align}
in which both the CSI $\boldsymbol{H}_k$ and the observation matrix $\boldsymbol{Y}_k$ are provided to the receiver. 
In our implementation, the CSI is first converted to a real-valued vector of length $2 N_R N_T$ and then concatenated to the observation matrix, which is also converted to a real-valued vector. 
Finally, an estimate of the transmitted message vector $\hat{\vect{m}}_k$ can be obtained based on $\argmax_m[\boldsymbol{q}_k]_m$ by inverting the one-hot encoding. 

%Note that while the transmitter does not have access to \emph{instantaneous} CSI in the learning process, it can obtain knowledge of the CSI distribution $p(\boldsymbol{h})$, i.e., \emph{statistical} CSI. 

\subsection{Numerical Results and Discussion}

\renewcommand{\arraystretch}{1.0}
\begin{table}[t]
\setlength{\tabcolsep}{0.6em}
\scriptsize
\centering
\vspace{0.15cm}
\caption{NN parameters for (i) open-loop MIMO, (ii) closed-loop MIMO, (iii) MIMO broadcast, and (iv) interference channel}
\begin{tabular}{c|c|ccc|ccc}
\toprule
& & \multicolumn{3}{c}{transmitter(s) $f_\tau$/$f_{\tau_i}$}   & \multicolumn{3}{|c}{receiver(s) $f_\rho$/$f_{\rho_i}$} \\
 \midrule
& layer    & input  & hidden& output    & input & hidden     & output  \\ 
\midrule
%open-loop:\\
\multirow{3}{*}{(i)}&\#~of layers    & -  & $3$ & -    & - & $3$     & -  \\ 
&\#~of neurons   & $M^2$ & 64   & $8$      & $12$     & $512$      & $M^2$          \\ 
&act.~function & - & ReLU & linear & -     & ReLU    & softmax    \\ 
\midrule
%closed-loop:\\
\multirow{3}{*}{(ii)}&\#~of layers    & - & $3$  & -    & - & $3$     & -  \\ 
&\#~of neurons   & $M+8$ & $1024$   & $8$      & $16$     & $1024$      & $M$          \\ 
&act.~function & - & ReLU & linear & -     & ReLU    & softmax    \\ 
\midrule
%MIMO broadcast:\\
\multirow{3}{*}{(iii)}&\#~of layers    & - & $3$  & -    & - & $3$     & -  \\ 
&\#~of neurons   & $M^2+8$ & $512$   & $8$      & $6$     & $256$      & $M$          \\ 
&act.~function & - & ReLU & linear & -     & ReLU    & softmax    \\ 
\midrule
%interference channel:\\
\multirow{3}{*}{(iv)}&\#~of layers    & - & $1$  & -    & - & $1$     & -  \\ 
&\#~of neurons   & $M$ & $256$   & $8$      & $8$     & $256$      & $M$          \\ 
&act.~function & - & ReLU & linear & -     & ReLU    & softmax    \\ 
\bottomrule
\end{tabular}
\label{tab:network_parameters}
\end{table}

%\begin{comment}
\renewcommand{\arraystretch}{1.0}
\begin{table}[t]
\setlength{\tabcolsep}{0.6em}
\scriptsize
\centering
\vspace{0.15cm}
\caption{Training parameters for (i) open-loop MIMO, (ii) closed-loop MIMO, (iii) MIMO broadcast, and (iv) interference channel}
\begin{tabular}{c|cccc}
\toprule
& (i) & (ii) & (iii) & (iv) \\ 
\midrule
%perf.~metric & $\text{Pr}\{\hat{\vect{m}}_k \neq \vect{m}_k\}$ & $\text{Pr}\{\hat{m}_k \neq m_k\}$ &$\text{Pr}\{\hat{m}_{k,i} \neq m_{k,i} \}$ &$\text{Pr}\{\hat{m}_{k,1} \neq m_{k,1} \}$ \\ 
optimizer & Adam & Adam & Adam & Adam \\ 
learning rate & $10^{-3}$ & $10^{-3}$ & $10^{-3}$ & $10^{-3}$ \\
batch size $B$ & $65536$ & $10240$ & $10240$ & $65536$ \\
grad. steps G & $2\times 10^4$ & $2\times 10^6$ & $4\times 10^5$ & $5 \times 10^4$ \\
training SNR & \rr{$15\,$dB ($M=4$)}{$18\,$dB ($M=16$)} & \rr{$5\,$dB, $10\,$dB}{$15\,$dB, $12\,$dB} & \rr{$12\,$dB, $15\,$dB}{$18\,$dB, $20\,$dB} & \rr{$11\,$dB}{($E_b/N_0$)} \\
%channel real.& $3\,200\,000$ & $10$ & $10$ & $10$ \\
\bottomrule
%$^*$ 
\end{tabular}
\label{tab:hyperparameters}
\end{table}
%\end{comment}

The channel is assumed to be Rayleigh fading, i.e., 
$\mathrm{vec}(\boldsymbol{H}_k) \sim \mathcal{CN}(\boldsymbol{0},\boldsymbol{I}_{N_R N_T})$. 
The system performance is measured in terms of the $\text{BLER} = \text{Pr}\{\hat{\vect{m}}_k \neq \vect{m}_k\}$ as a function of the average $\text{SNR}=P_T/(N_T N_0)$.
We use the parameters $N_T=2$, $N_R=1$, $N_B=2$, $L=2$, and $M\in \{ 4,16\}$. 
In this paper, all AEs are implemented as multi-layer fully-connected NNs, where the rectified linear unit (ReLU) is chosen as the activation function. 
% Both the number of hidden layers and the number of neurons per layers were optimized and the resulting NN parameters for all scenarios are summarized in Table~\ref{tab:network_parameters}.
\RevB{To optimize the number of hidden layers and the number of neurons per layer, several AEs with different sizes are trained for each of the considered scenarios, and we then choose the AEs with the best performance.} The resulting NN parameters for all scenarios are summarized in Table~\ref{tab:network_parameters}.\footnote{\RevB{We remark that the NN parameters used in this paper are not guaranteed to be fully optimal, though further optimization of the NN parameters are not expected to improve the AE performance significantly.}}
\RevB{Moreover, all AEs are trained by using the Adam optimizer \cite{Kingma2015}, where the learning rate, batch size, and the number of gradient steps are summarized in Table~\ref{tab:hyperparameters}. In particular, i.i.d.\,training samples are randomly generated in each training iteration, and the total number of samples used for training each of the considered AEs is $B\times G$, where $G$ denotes the number of gradient steps. For the performance evaluation,  independent testing data are continuously generated until at least $5\times 10^5$ errors are counted for each considered SNR.}

% \RevB{Particularly, for the open-loop case, the AEs are trained with $2\times 10^4$ gradient steps, where in each gradient step a minibatch of 65536 independent random messages and channel realizations are sampled from the corresponding message and channel distribution. Therefore, a total number of $2\times 10^4 \times 65536$ training data are used to trained each of the open-loop AE. For the testing, to avoid leakage of training data into the test set, we generate random independent testing data in the same way as we generate the training data. } 

\begin{figure}[t]
    \centering
    \includegraphics[width=\columnwidth]{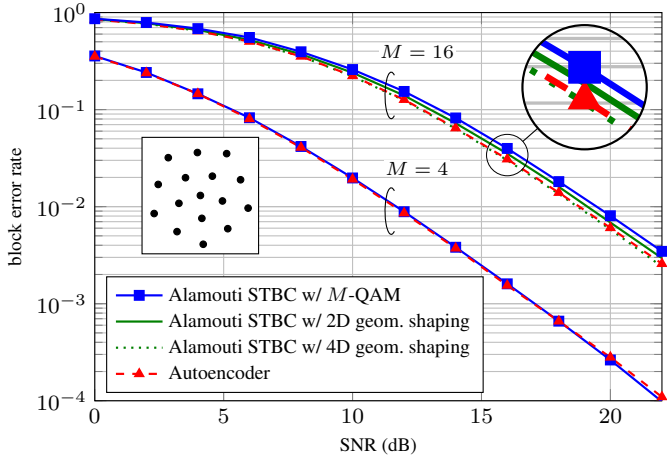}
    %\vspace{-0.1cm}
    \caption{BLER of the open-loop MIMO AE and the baseline scheme consisting of standard $M$-QAM signal constellations, an Alamouti STBC, and a maximum-likelihood receiver. The improved baselines for $M=16$ use geometrically-shaped signal constellations in two and four dimensions, respectively.}% Training is done at SNR$=15\text{dB}$ for $M=4$ and at SNR$=18\text{dB}$ for $M=16$ .}
    \label{fig:alamouti}
\end{figure}

\begin{figure}[t]
    \centering
    \subfloat[]{\includegraphics[width=4.0cm]{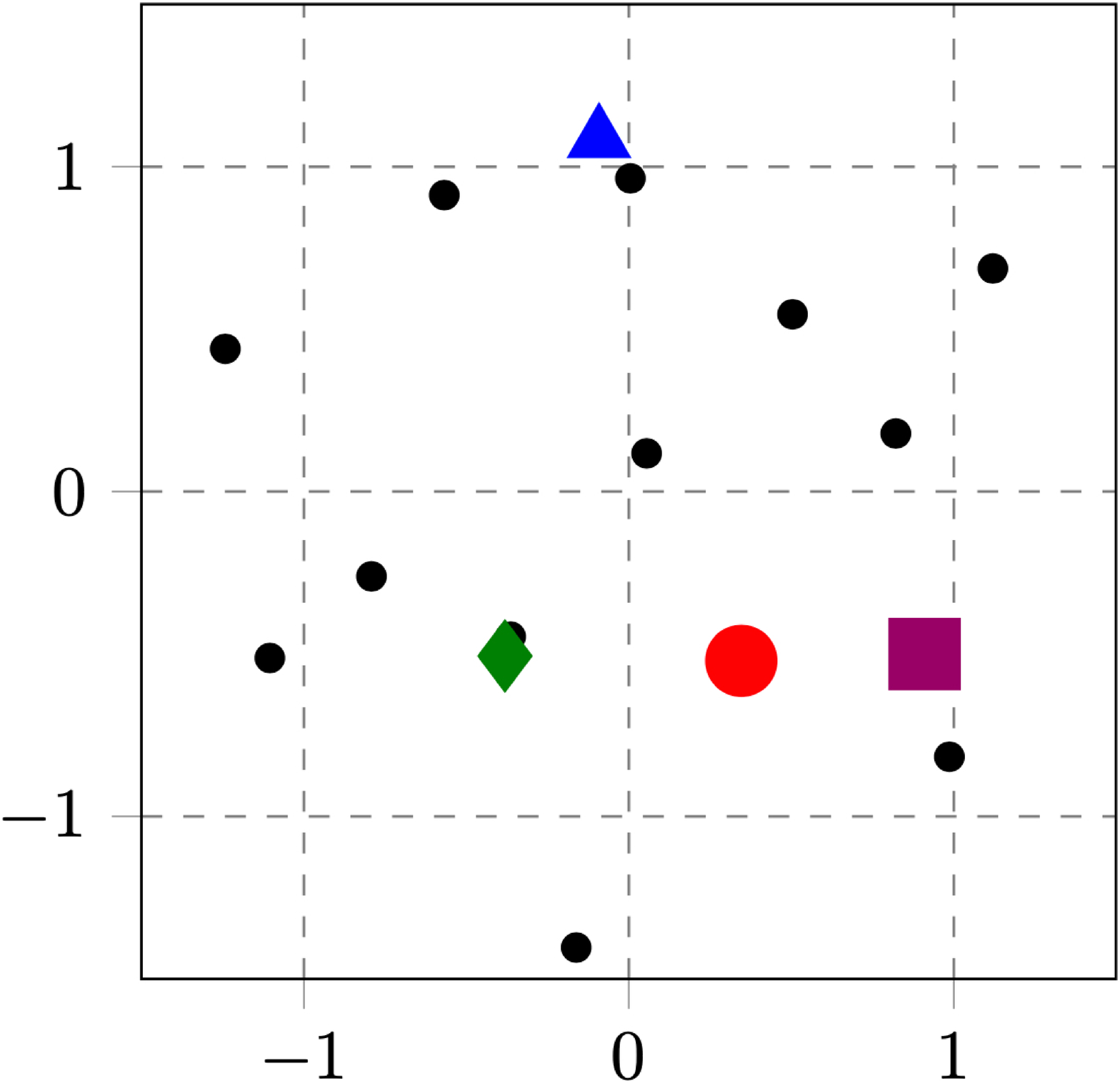}} $\quad$
    \subfloat[]{\includegraphics[width=4.0cm]{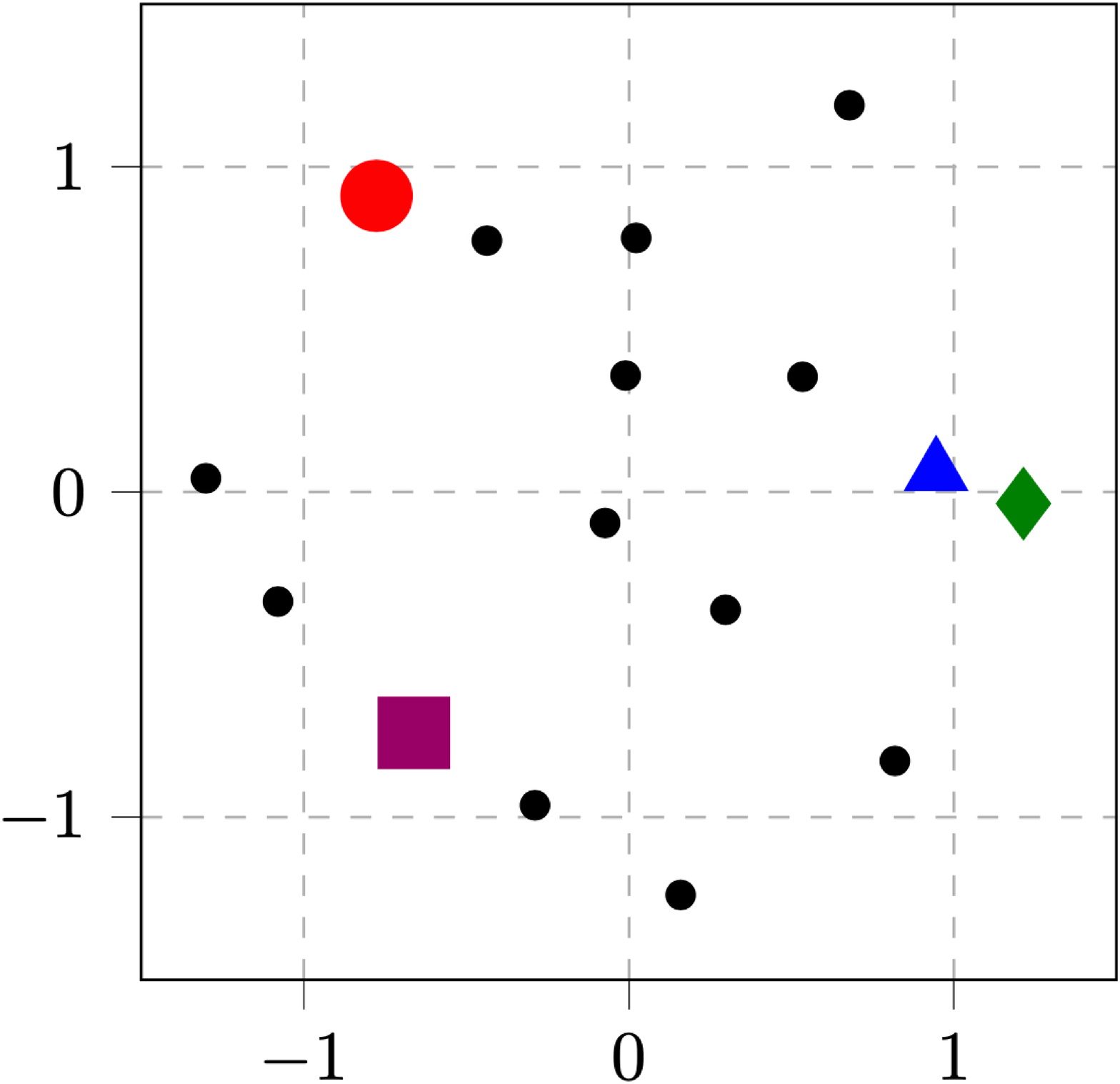}}\\
    \subfloat[]{\includegraphics[width=4.0cm]{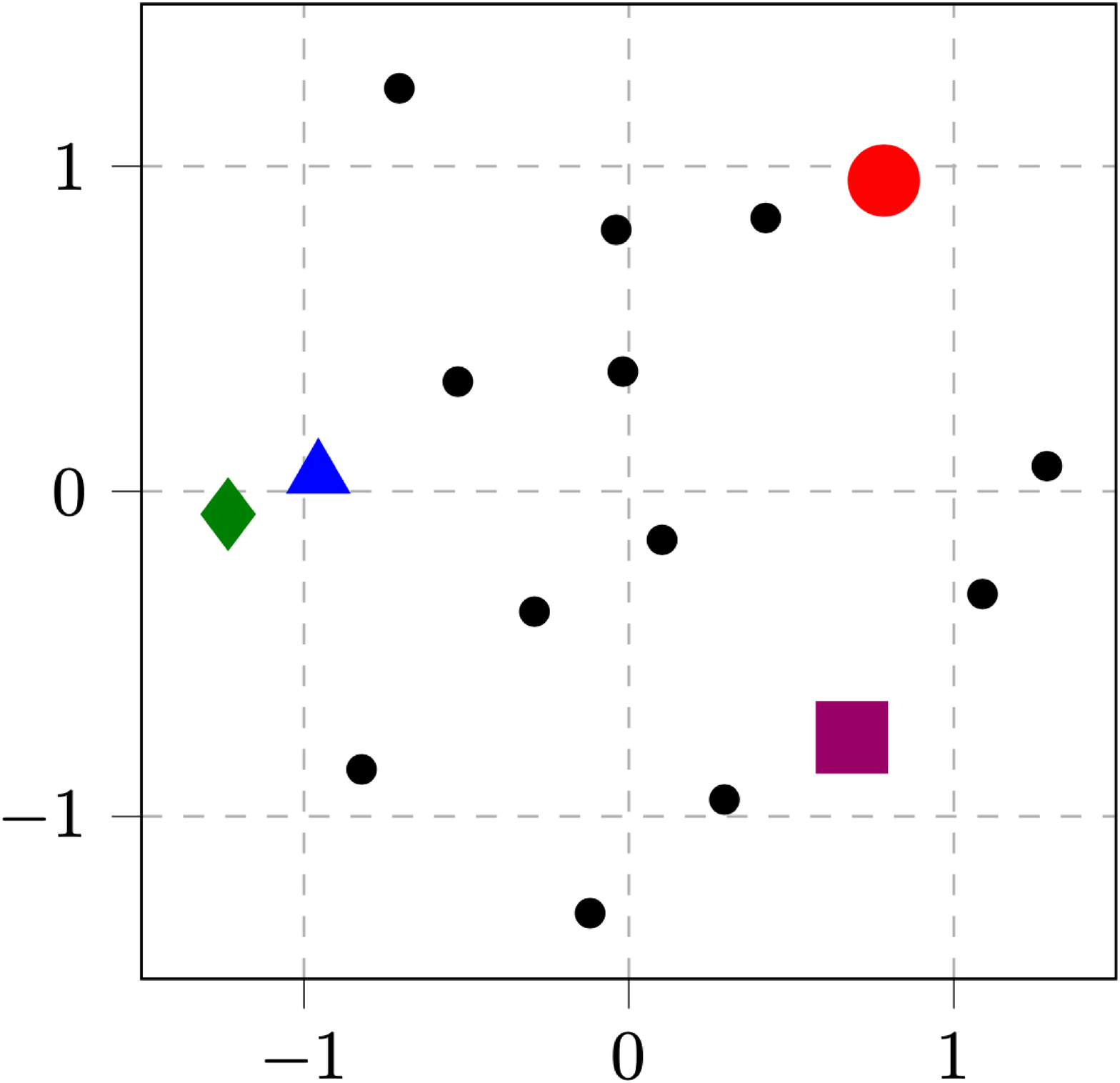}} $\quad$
    \subfloat[]{\includegraphics[width=4.0cm]{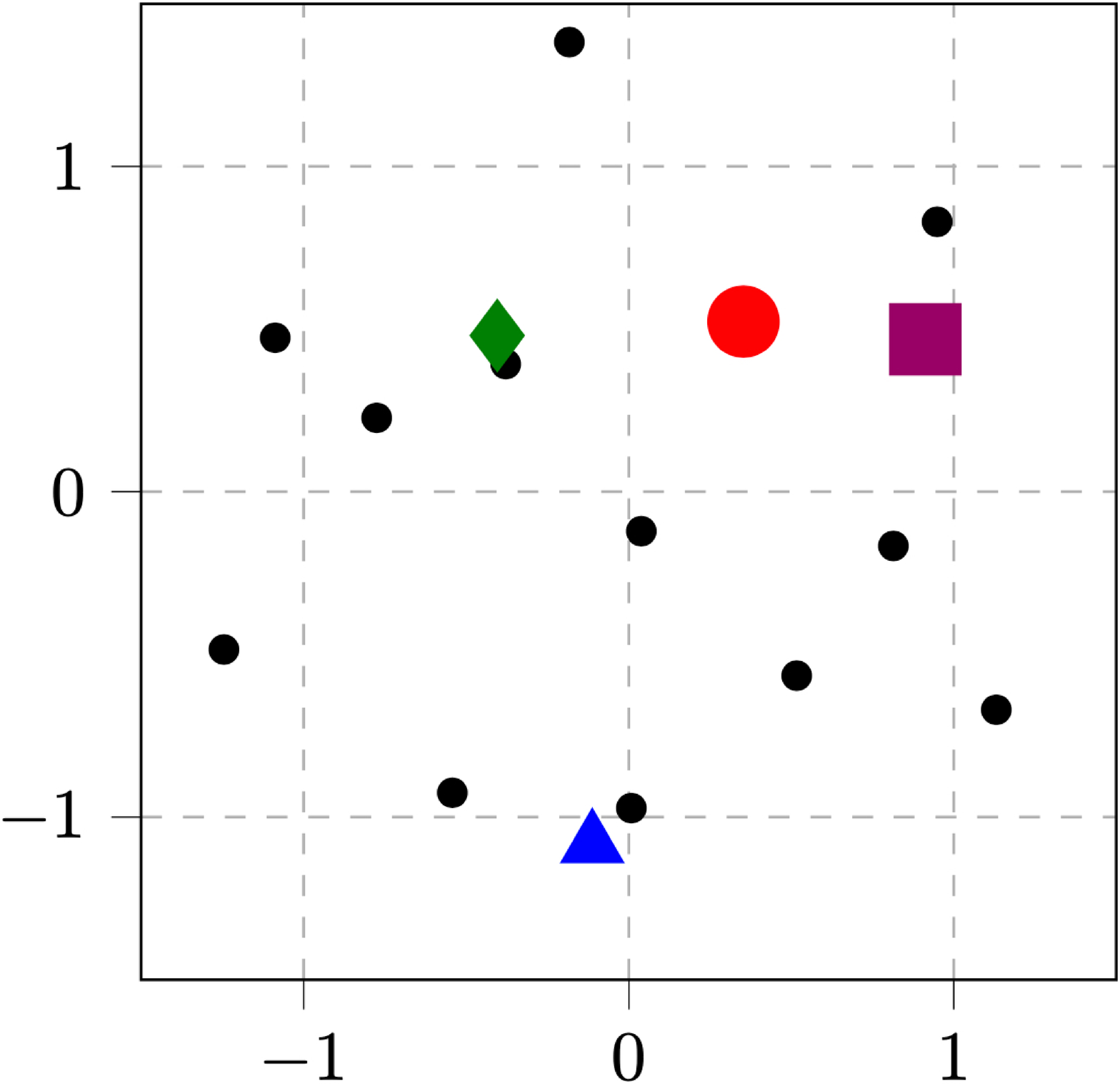}}
    \caption{Learned transmitted symbols of the open-loop MIMO AE for $M=4$. (a) first antenna at time slot $1$, (b) second antenna at time slot $1$, (c) first antenna at time slot $2$ and (d) second antenna at time slot $2$. Constellation points for $4$ out of $16$ messages are highlighted with colored markers.} 
    \label{fig:learnt_const}
\end{figure}

Fig.~\ref{fig:alamouti} shows the achieved BLER over a range of SNRs (red triangles). 
As a reference, the performance of the baseline Alamouti scheme with $M$-QAM constellations is also shown (blue squares). 
For $M=4$, the AE achieves very similar performance to the baseline scheme, indicating that the combination of a QPSK constellation and Alamouti STBC is near-optimal in this case. 
For $M=16$, the AE outperforms the baseline scheme at medium-to-high SNRs by about $0.6\,$dB when standard $16$-QAM is used as the signal constellation. 
In order to improve the baseline for $M=16$, we also used two geometrically-shaped (GS) signal constellations, which were obtained by training a standard AE over an AWGN channel.\footnote{\RevB{To obtain each of the GS signal constellations, we trained several pairs of AEs over the AWGN channel at different SNRs and then chose the one with the best performance.}}
The first constellation has $16$ points in two dimensions and is shown in the inset figure in Fig.~\ref{fig:alamouti}.
Its performance sits approximately halfway between the AE and the STBC with $16$-QAM. 
The second constellation has $M^2 = 256$ points in four dimensions. 
In this case, the constellation is first mapped to $\vect{s}_k = [s_{k,1}, s_{k,2}]^\transpose$, after which the standard Alamouti code can be applied.
When this four-dimensional constellation is used instead, the baseline scheme has essentially the same performance as the AE-based approach.\footnote{Note that the four-dimensional format does not necessarily admit a low-complexity detection separately based on $\tilde{s}_{k,1}$ and $\tilde{s}_{k,2}$. In our implementation, the decoding is instead performed using \eqref{eq:open_loop_ml}, where the optimization is over all $256$ constellation points.}

The results presented here do not confirm the preliminary results presented in \cite{OShea2017Deep}, where it was found that the AE outperforms the Alamouti scheme at high SNR. 
One potential reason for this discrepancy could be the different power normalization that is used in \cite{OShea2017Deep} after applying the channel matrix (cf.~\cite[Eq.~(2)]{OShea2017Deep}).
Instead, our results indicate that the AE learns to perform a joint optimization over the signal constellation and STBC, where the AE recovers the well-known Alamouti code for the considered scenario. 
To further support this observation, Fig.~\ref{fig:learnt_const} visualizes the learned transmitted symbols for $M=4$ after applying a $2$-dimensional rotation to the symbols. 
Particularly, the constellation points for $4$ out of $M^L=16$ individual messages are highlighted by different markers. 
From these plots, one can observe that the learned constellation follows a very similar pattern as the Alamouti scheme, in the sense that the symbols in the upper left subplot (a) are symmetric with respect to the ones in the lower right (d) subplot along the x-axis, while the symbols in the upper right subplot (b) are symmetric with respect to the ones in the lower left subplot (c) along the y-axis.

\section{Closed-Loop MIMO}
\label{sec:closed-loop}
In closed-loop MIMO systems, the CSI is available at both the transmitter and receiver. The most common approach in this case is SVD-based transmission, which we describe in the next subsection.
\subsection{Background and Baseline Schemes}
%The block fading duration is irrelevant, but should be long enough to allow feedback and use of the CSI $\boldsymbol{H}_k$. 
Both the transmitter and receiver compute the SVD
\begin{align}
\label{eq:svd}
   \boldsymbol{H}_k = \boldsymbol{U}_k\boldsymbol{\Sigma}_k\boldsymbol{V}_k^\Ctranspose,  
\end{align}
where $\boldsymbol{\Sigma}_k=\text{diag}[\sigma_{k,1},\ldots, \sigma_{k,R_H}]$,  $\sigma_{k,1}\ge \sigma_{k,2} \ge \cdots \ge \sigma_{k,R_H}>0$ and $R_H$ is the rank of $\boldsymbol{H}_k$. 
Correspondingly, $\boldsymbol{U}_k\in \mathbb{C}^{N_R\times R_H}$ and $\boldsymbol{V}_k\in \mathbb{C}^{N_T\times R_H}$ are truncated unitary matrices.
For each singular value $\sigma_{k,i}$, the transmitter chooses a constellation $\Omega_{i}$ from a set of available constellations, as well as a transmit power $P_{T,i}\ge 0$. This selection can be based on the total BLER according to
\begin{subequations}
\label{eq:svd_baseline}
\begin{align}
    \underset{\Omega_{i},P_{T,i}}{\text{minimize}}~~~ & \textstyle{ 1- \prod_{i=1}^{R_H}}(1-P_e(\Omega_{i},\gamma_i))\\
    \text{s.t.}~~~& \textstyle{\prod_{i=1}^{R_H}} |\Omega_{k,i}|=M,\\
     & \textstyle{\sum_{i=1}^{R_H}} P_{T,i}\le P_T,\\
     & \gamma_i = \frac{\sigma^2_{k,i}P_{T,i}}{N_0},
\end{align}
\end{subequations}
where $P_e(\Omega,\gamma)$ is the symbol error probability of constellation $\Omega$ under the specific SNR $\gamma$.  
Hence, the rate is fixed to $r=\log_2(M)$. 
The corresponding symbol vector $\boldsymbol{s}_k = [s_{k,0}, s_{k,1},\cdots, s_{k,R_H}]^\transpose$ is precoded by $\boldsymbol{V}_k$, so that $\boldsymbol{x}_k=\boldsymbol{V}_k\boldsymbol{s}_k$ is sent over the channel, where $\mathbb{E}\{\Vert \boldsymbol{x}_k\Vert ^2 | \mat{H}_k\} = P_T$. 
The receiver observes $\boldsymbol{y}_k  = \boldsymbol{H}_k\boldsymbol{x}_k + \boldsymbol{n}_k$ and applies the combiner $\boldsymbol{U}^\Ctranspose_k$, leading to the observation
\begin{align}
\label{eq:observation}
    \boldsymbol{\hat{y}}_k=\boldsymbol{U}^\Ctranspose_k\boldsymbol{H}_k\boldsymbol{V}_k\boldsymbol{s}_k + \boldsymbol{U}^\Ctranspose_k\boldsymbol{n}_k = \boldsymbol{\Sigma}_k\boldsymbol{s}_k+ \boldsymbol{U}^\mathsf{H}\boldsymbol{n}_k.
\end{align}
%where $ \boldsymbol{U}^\mathsf{H}\boldsymbol{n}_k$ has the same distribution as $\boldsymbol{n}_k$. <- only if R_H = N_R
Maximum-likelihood recovery of the transmitted messages is straightforward since $\boldsymbol{\Sigma}_k$ is a diagonal matrix.

\subsection{Autoencoder Design and Training}
 \begin{figure}[t]
     \centering
     \includegraphics[width=8.7cm]{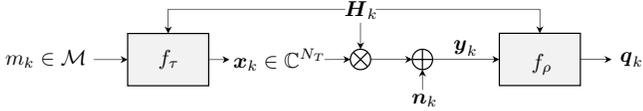}
     %\vspace{-0.1cm}
     \caption{Closed-loop MIMO AE, in which both the transmitter and receiver have access to CSI.  }
     \label{fig:closed-loopSYSTEM}
 \end{figure}
The AE for a closed-loop MIMO system is implemented as shown in Fig.~\ref{fig:closed-loopSYSTEM}. 
To provide the transmitter with CSI, the corresponding NN is of the form $f_{\tau}:\mathcal{M}\times \mathbb{C}^{N_R\times N_T} \to \mathbb{C}^{N_T\times1} $, yielding complex transmit vectors $\boldsymbol{x}_k=f_\tau(m_k, \boldsymbol{H}_k)$.
As before, a one-hot encoding is used to map the message $m_k$ to a vector of length $M$, which is then concatenated with the vectorized real and imaginary parts of the channel matrix. 
To enforce the power constraint $\mathbb{E}\{\Vert \boldsymbol{x}_{k} \Vert ^2 | \mat{H}_k\} = P_T$, the normalization layer is defined by
\begin{align}
    \label{eq:normalization_mod}
    \vect{x}_k = \frac{\tilde{\vect{x}}_k \sqrt{P_T}}{\sqrt{\frac{1}{M}\sum_{i=1}^M 
    \| 
    %\tilde{\vect{x}}_{k'} 
    \tilde{f}_\tau(i, \mat{H}_k)
    \|^2}},
\end{align}
where $\tilde{\vect{x}}_k = \tilde{f}_\tau(m_k, \mat{H}_k)$ is the unnormalized NN output. 
Thus, even though $\vect{x}_k$ is a function of the (random) channel realization $\vect{H}_k$, the expectation $\mathbb{E}\{\Vert \boldsymbol{x}_{k} \Vert ^2  | \mat{H}_k\}$ is performed only over the messages. 
This ensures that the AE output is always normalized, even if the actual channel distribution deviates from the distribution used for training. 

Finally, the receiver $f_{\rho}: \mathbb{C}^{N_R\times 1} \times \mathbb{C}^{N_R\times N_T} \to [0,1]^{M}$ observes $\boldsymbol{y}_k = \boldsymbol{H}_k \boldsymbol{x}_k + \boldsymbol{n}_k$ and, similarly to the open-loop MIMO case, the transmitted message is estimated as $\hat{m}_k=\argmax_m[\boldsymbol{q}_k]_m$, where $\boldsymbol{q}_k=f_\rho(\boldsymbol{y}_{k}, \boldsymbol{H}_k)$ is a probability vector obtained in the same way as in \eqref{eq:NN_decoder}.

To generate a minibatch (of size $B$) for the Monte Carlo approximation of the cross-entropy loss \eqref{eq:ce_loss}, we first randomly generate $B/M$ i.i.d.~channel realizations. 
Then, for each channel realization all distinct $M$ messages are assumed to be transmitted. 
Compared to the approach of generating random messages and channel realizations for each data sample, this has the advantage that the normalization factor in the denominator of \eqref{eq:normalization_mod} can be applied to $M$ messages at once and does not need to be computed for every data sample in the batch. \RevB{The same approach is used to generate the testing data.}

\subsection{Numerical Results and Discussion}
\label{sec:closed-loop_results}

We consider Rayleigh fading and use the parameters $N_T=2, N_R=2$, and $M=16$, corresponding to rate $r=4$. 
In this case, $\text{BLER} = \text{Pr}\{\hat{m}_k \neq m_k \} $ and $\text{SNR}=P_T/(N_T N_0)$. 
The NN and training parameters are shown in Tabs.~\ref{tab:network_parameters} and \ref{tab:hyperparameters}, respectively. 
Compared to the open-loop case, we noticed that more data samples are required for converging to a good solution. 
Moreover, varying the SNR throughout the training was found to improve performance, \RevB{which was not observed for the open-loop case}. In particular, we train the AE consecutively at $5\,$dB $\to$ $10\,$dB $\to$ $15\,$dB $\to$ $12\,$dB, where each SNR is kept fixed for  $5\times 10^5$ iterations, giving $2 \times 10^6$ training iterations in total.

\begin{figure}[t]
    \centering
    \includegraphics[width=\columnwidth]{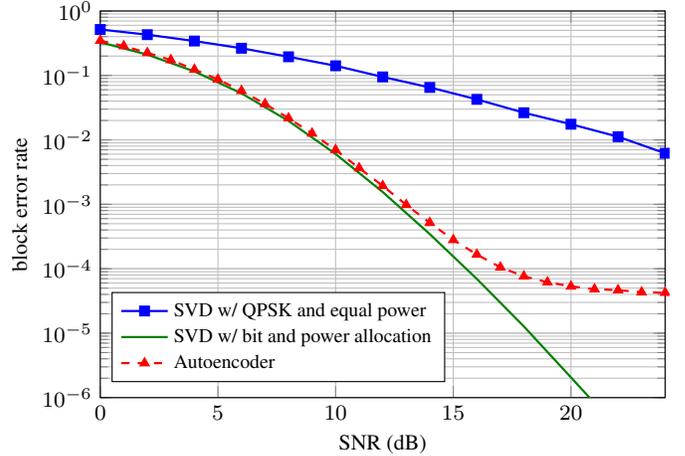}
    %\vspace{-0.1cm}
    \caption{BLER of the closed-loop MIMO AE for $M=16$ and the baseline scheme consisting of a QPSK constellation, SVD-based signal processing, and a maximum-likelihood receiver. The improved baseline uses bit and power allocation. }%Training is done for a fixed SNR$=15\text{dB}$.}
    \label{fig:svd-ser}
\end{figure}
The BLER achieved by the trained AE is shown in Fig.~\ref{fig:svd-ser} (red triangles). 
As a baseline, we simulate the performance of the SVD-based approach, in which the $2\times2$ MIMO channel is parallelized into two subchannels. 
We first consider the same baseline as in \cite{OShea2017Deep}, where equal power is used at each antenna and both streams use QPSK modulation (blue squares in Fig.~\ref{fig:svd-ser}). 
Similarly to what was observed in \cite{OShea2017Deep}, the AE achieves significantly better performance than the SVD-based approach with QPSK and equal power allocation. 
However, depending on the channel realization, the two individual subchannels will have different link quality, and bit and power allocation are usually used to improve the overall system performance. 
To that end, an improved baseline scheme was simulated by solving \eqref{eq:svd_baseline} using exhaustive search assuming that the set of available signal constellations is BPSK, QPSK, 
{\tt c2\_8} \cite{codes.se}, and {\tt c2\_16} \cite{codes.se}, where the latter two are geometrically-shaped 2-dimensional constellations with $8$ and $16$ points, respectively.\footnote{Rectangular $8$-QAM and $16$-QAM were used in \cite{Song2020Globecom} which give slightly worse performance.}
As can be seen in Fig.~\ref{fig:svd-ser}, this improved baseline provides slightly better BLER than the AE at low SNR. At high SNR, the baseline significantly outperforms the AE, which exhibits an error floor between BLERs of $10^{-4}$ and $10^{-5}$. 
This error floor is caused by the fact that the transmitter NN takes the channel as an input. 
Indeed, depending on the particular channel realization, we noticed that the transmitter NN sometimes produces a signal constellation that has very poor performance. 
While such outliers are rare, they dominate the average performance at very high SNR.
We also note that the error floor can be lowered by retraining the AE at a higher SNR, but this may come at the expense of some performance loss in the low SNR regime. 

The above results indicate that the closed-loop MIMO AE learns to implicitly perform a combination of GS, bit allocation, and power allocation.
In fact, it is insightful to further examine the performance of the trained AE assuming that the singular values of the channel matrix remain constant. 
To evaluate the AE, channel matrices can be generated by using random unitary matrices for $\mat{U}_k$ and $\mat{V}_k$ in \eqref{eq:svd}. 
\begin{figure}[t]
    \centering
    %[mode=build]
    % \includestandalone{tikz/fixed_svs}
     \includegraphics[width=\columnwidth]{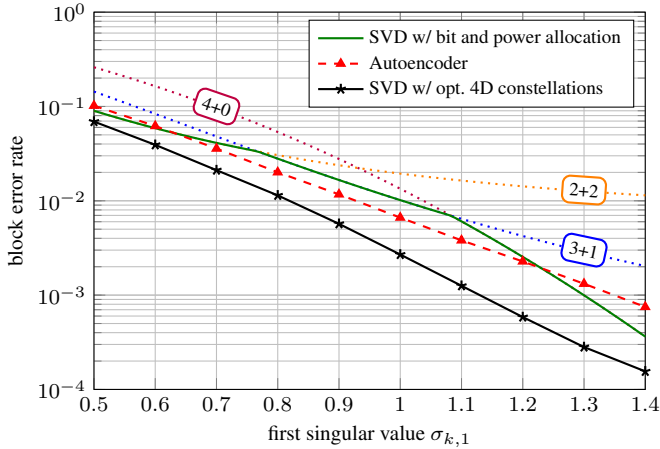}
    %\vspace{-0.1cm}
    \caption{BLER for the closed-loop MIMO system at $\text{SNR} = 12$ dB over channel matrices with fixed singular values (without retraining). The value of the second singular value is $\sigma_{k,2} = 0.5$. The dotted lines correspond to the SVD-based baseline with fixed bit allocation for the two parallel channels. }%
    \label{fig:fixed_evs}
\end{figure}
Fig.~\ref{fig:fixed_evs} shows the resulting AE performance (without any retraining) as a function of the first singular value $\sigma_{k,1}$, where the second singular value is $\sigma_{k,2} = 0.5$. 
It can be seen that the AE actually outperforms the SVD-based baseline in the range $0.7 \leq \sigma_{k,1} \leq 1.2$, even when bit and power allocation are used. 

To further improve the SVD-based baseline, we also trained a standard AE directly for the observation model \eqref{eq:observation} assuming fixed singular values, according to the methodology in Sec.~\ref{sec:siso_ae}. 
This essentially provides optimized $4$-dimensional signal constellations over two parallel AWGN channels with different (but fixed) SNRs. 
The resulting performance is shown in Fig.~\ref{fig:fixed_evs} by the black markers, where the optimization is performed separately for each $(\sigma_{k,1}, \sigma_{k,2})$ with $\sigma_{k,1} \in \{ 0.5, 0.6, \ldots, 1.4\}$ and $\sigma_{k,2} = 0.5$. 
This approach provides the best performance among all considered schemes. However, it has the downside that a separate optimization is required for each pair of singular values.  
Nonetheless, this approach does provide additional insight into why the AE can outperform the SVD-based baseline with bit and power allocation for some channel configurations. 
In particular, the suboptimality of the latter scheme stems from the fact that the two parallel subchannels are treated independently, whereas the AE treats all available signal dimensions in a joint manner.

\section{MIMO Broadcast Channel}
\label{sec:mu-mimo}

In this section, we consider a downlink MIMO system where one transmitter with $N_T$ antennas broadcasts messages to $N_R$ receivers each with one antenna, where $N_T\geq N_R$. 
This scenario is sometimes also referred to as the multiple-input single-output broadcast channel \cite{Wiesel2008}.

\subsection{Background and Baseline Schemes}

It is assumed that local CSI $\boldsymbol{h}^{\mathsf{T}}_{k,i}\in \mathbb{C}^{1\times N_T}$ is available at each receiver $i=1,\ldots,N_R$, whereas the transmitter has knowledge of the full CSI $\boldsymbol{H}_k=[\boldsymbol{h}_{k,1},\ldots,\boldsymbol{h}_{k,N_R} ]^\mathsf{T}$.
%Similarly to the closed-loop MIMO case in the previous section, the block fading duration is irrelevant, but should be long enough to allow feedback and use of the CSI. 
To manage the interference among different users, various algorithms have been proposed \cite{Joham2005, Hochwald2005, Gesbert2007, Wiesel2008}. 
In this paper, we consider both a linear precoding scheme referred to as transmitter ZF and a nonlinear vector-perturbation scheme. Both schemes are described next.
 
For linear precoding, the transmitter first maps $N_R$ messages $m_{k,1},\ldots,m_{k, N_R}$ to symbols $s_{k,1},\ldots,s_{k,N_R}$. A precoding matrix $\mat{W}_k \in \mathbb{C}^{N_T \times N_R}$ is then used to encode $\boldsymbol{s}_k \define [s_{k,1},\cdots,s_{k,N_R}]^\mathsf{T}$ according to $\tilde{\vect{x}}_k = \mat{W}_k\boldsymbol{s}_k$.
Afterwards, a normalized version $ \boldsymbol{x}_k = \alpha \tilde{\vect{x}}_k$ is sent over the channel to ensure that $\mathbb{E}\{\Vert \boldsymbol{x}_k\Vert ^2 | \mat{H}_k \} = P_T$, where $\alpha \define \sqrt{P_T / \mathbb{E}\{\|\tilde{\vect{x}}_k\|^2 | \mat{H}_k \}}$ and the expectation is with respect to the messages of all users. 
The scaling factor $\alpha$ is assumed to be known to all receivers. 
Each of the symbols is assumed to carry $\log_2(M)$ bits and, consequently, the sum-rate of the system is $r = N_R\log_2(M)$.
The precoding matrix is of the form
\begin{align}
\label{eq:precoding_matrix}
    \mat{W}_k  =\boldsymbol{H}_k^\Ctranspose(\boldsymbol{H}_k\boldsymbol{H}_k^\Ctranspose + \beta \mat{I}_{N_R})^{-1}, 
\end{align}
where $\beta \in \mathbb{R}$ is a regularization parameter. 
For ZF, we have $\beta = 0$ and $\mat{W}_k$ then corresponds to the pseudoinverse of the channel matrix. 
In this case, each user $i$ observes  $y_{k,i} = \boldsymbol{h}^{\mathsf{T}}_{k,i} \boldsymbol{x}_k + n_{k,i}=\alpha s_{k,i} + n_{k,i},$ from which $s_{k,i}$ can be recovered with low-complexity maximum-likelihood detection.

We also consider the nonlinear precoder proposed in \cite{Hochwald2005}. 
Compared to ZF, the transmitter computes the unnormalized transmit vector according to $ \tilde{\boldsymbol{x}}_k= \boldsymbol{W}_k(\boldsymbol{s}_k + \vect{p}_k^*)$, where $\mat{W}_k$ is again defined by \eqref{eq:precoding_matrix} (potentially with $\beta > 0$),
\begin{align}
\label{eq:perturbation}
    \vect{p}_k^* = \argmin_{\vect{p'} \in A \mathbb{C}\mathbb{Z}^{N_R}} \| \mat{W}_k(\vect{s}_k + \vect{p}') \|^2
\end{align}
is a perturbation vector from the scaled complex integer lattice $\mathbb{C}\mathbb{Z}^{N_R} \define \{\vect{x}+\jmath \vect{y} : \vect{x},\vect{y} \in \mathbb{Z}^{N_R} \}$, and the scaling factor $A$ depends on the modulation format. 
Each receiver first applies a modulo operation $z_{k,i} = \text{cmod}_A(y_{k,i}/\alpha) \in \mathbb{C}$, where $\text{cmod}_A(x_r + \jmath x_i) = \text{mod}_A(x_r) + \jmath \text{mod}_A(x_i)$ and
\begin{align}
    \text{mod}_A(x) \define x - A \lfloor (x + A/2)/A \rfloor
\end{align}
for $x_r, x_i, x \in \mathbb{R}$. 
Afterwards, one can again apply low-complexity maximum-likelihood detection based on $z_{k,i}$. 

%closest-point lattice search \cite{Fincke1985}
%dirty paper coding \cite{Costa1983}. 

\subsection{Autoencoder Design and Training}
%\subsection{MIMO Broadcast AE}

\begin{figure}[t]
    \centering
    \includegraphics[width=8.7cm]{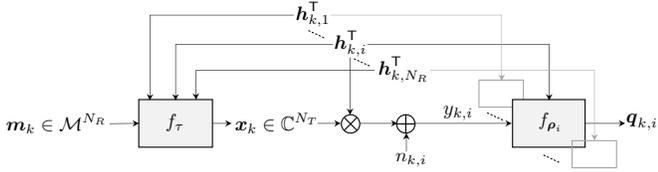}
    %\vspace{-0.1cm}
    \caption{MIMO broadcast AE, in which the transmitter encodes messages for the individual users, based on full CSI, while each user observes only a local measurement and local CSI. }
    \label{fig:mu-mimoSYSTEM}
\end{figure}

The proposed AE implementation for the MIMO broadcast channel is visualized in Fig.~\ref{fig:mu-mimoSYSTEM}. 
The transmitter $f_\tau: \mathcal{M}^{N_R} \times  \mathbb{C}^{N_R\times N_T} \to \mathbb{C}^{N_T\times 1}$ maps individual messages $m_{k,i}\in \mathcal{M}$ for each user $i=1,\cdots, N_R$ to $N_T$ complex symbols according to $\boldsymbol{x}_k = f_\tau(\boldsymbol{m}_k, \boldsymbol{H}_k)$, where $\boldsymbol{m}_k=[m_{k,1},\cdots,m_{k, N_R}]^{\mathsf{T}}$. One-hot encoding of $\boldsymbol{m}_k$ to a vector of length $M^{N_R}$ is applied.
The power constraint $\mathbb{E}\{\Vert \boldsymbol{x}_k\Vert ^2 | \mat{H}_k \} = P_T$ is enforced through a normalization layer similar to \eqref{eq:normalization_mod}, where the sum in the denominator runs over the messages of \emph{all} users. 

The $N_R$ receivers are implemented as $N_R$ individual NNs of the form $f_{\rho_i}:\mathbb{C} \times \mathbb{C}^{N_T} \to [0,1]^M$. 
In particular, each user $i$ observes $y_{k,i} = \boldsymbol{h}^\mathsf{T}_{k,i}\boldsymbol{x}_k + n_{k,i}$ and generates a probability vector $\boldsymbol{q}_{k,i}\in[0,1]^M$ according to 
\begin{align}
    \boldsymbol{q}_{k,i} = f_{\rho_i}(y_{k,i}, \boldsymbol{h}_{k,i}),
\end{align}
where the receiver network is provided with its observation $y_{k,i}$ as well as the local CSI $\boldsymbol{h}_{k,i}$. Then, the transmitted message for the $i$-th user is estimated as $\hat{m}_{k,i}=\argmax_m[\boldsymbol{q}_{k,i}]_m$.

In order to train the MIMO broadcast AE, the cross-entropy loss function defined in \eqref{eq:ce_loss} cannot be used directly, as we now have several receivers that need to be optimized. Instead, we apply a joint loss function 
\begin{align}
\label{eq: joint_ce}
%_{m_{k,i},\boldsymbol{y}_{k,i},\boldsymbol{h}_{k,i}}
 \mathcal{J}_{\text{CE}}(\tau, \rho_1, \cdots, \rho_{N_R}) =
 -\frac{1}{N_R}\sum_{i=1}^{N_R}\mathbb{E}\left\{  \log[f_{\rho_i}(\boldsymbol{y}_{k,i})]_{m_{k,i}} \right\},
\end{align}
which can again be optimized using the Adam optimizer.
    
\subsection{Numerical Results and Discussion}
\label{sec:mu-mimo_results}

As before, we consider Rayleigh fading and use the parameters $N_T=2$, $N_R=2$, and $M=4$, corresponding to a sum-rate $r=4$. 
Compared to the previous two cases, there are now three different NNs: one corresponding to the transmitter and two to the individual users, where the same NN architecture is used for both users, see Table~\ref{tab:network_parameters}. 
For simplicity, it is assumed that both receiver NNs share the same parameters, i.e., $\rho_1 = \rho_2$. 
Training is performed according to the parameters shown in Table~\ref{tab:hyperparameters}. 
Similarly to the closed-loop MIMO case, we found that it is beneficial to vary the SNR throughout the training.  

\begin{figure}[t]
    \centering
    % \includestandalone{tikz/mu_mimo_ser}
    \includegraphics[width=\columnwidth]{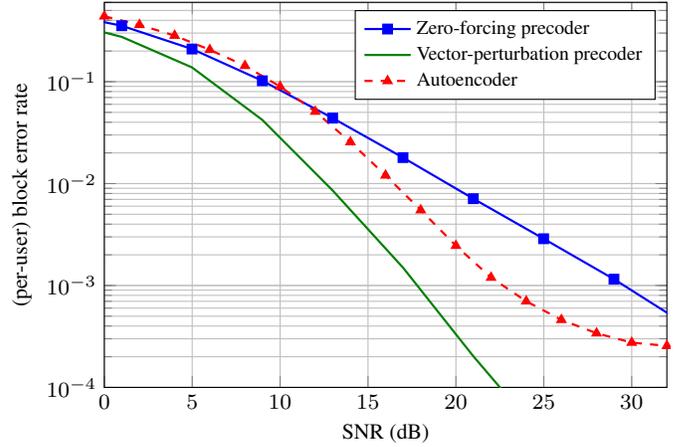}
    %\vspace{-0.1cm}
    \caption{BLER of the MIMO broadcast AE for $M=4$, $N_T = 2$ transmit antennas, and $N_R = 2$ users. Both baseline schemes use QPSK modulation. }
    \label{fig:mu-mimo}
\end{figure}

Fig.~\ref{fig:mu-mimo} shows the achieved BLER $\text{Pr}\{\hat{m}_{k,1} \neq m_{k,1} \}$ for the first user of the MIMO broadcast AE (red triangles) as a function of  $\text{SNR}=P_T/(N_T N_0)$, where the BLER for the second user is nearly identical and omitted. 
The performance of the ZF baseline approach with QPSK modulation, i.e., $s_{k,i} \in \{\pm 1 \pm \jmath\}/2$, is also shown (blue squares). 
It can be seen that the AE-based broadcast scheme achieves significantly better performance than the ZF approach for SNRs above $11\,$dB. 
For example, a gain of around $6\,$dB is achieved at a BLER of $10^{-3}$. 
Similarly to the closed-loop MIMO case, the AE exhibits an error floor which is affected by the training SNR and stems from the fact that the channel realization is taken is an input to the transmitter NN.

As a second baseline, we simulate the performance of the nonlinear vector-perturbation precoder. 
For QPSK modulation $s_{k,i} \in \{\pm 1 \pm \jmath\}/2$, the scaling factor is $A=2$ \cite{Hochwald2005} and \eqref{eq:perturbation} is solved approximately through exhaustive search, where the search space is restricted by replacing the entire integer lattice $\mathbb{Z}^{N_R}$ with a finite set $\{-5,-4,\ldots,4,5\}^{N_R}$.
The regularization parameter in \eqref{eq:precoding_matrix} is set to $\beta = \xi/\text{SNR}$, where $\xi=0.6$ was numerically optimized using a grid search. 
The resulting performance is shown by the solid green line in Fig.~\ref{fig:mu-mimo}. 
It can be seen that this nonlinear precoder outperforms the other two approaches for all SNRs.
Thus, our results show that the AE does not outperform a state-of-the-art baseline scheme for the considered MIMO broadcast scenario. 
However, we note that that the complexity associated with solving \eqref{eq:perturbation} is significant. 
Thus, the AE could potentially serve as a lower-complexity alternative, at the expense of some performance loss. 

In parallel to our work, a related AE-based approach for the MIMO broadcast channel was recently proposed in \cite{Pathapati2020}.
In this work, it is shown that the considered AE achieves significant performance advantages over Tomlinson--Harashima precoding \cite{Tomlinson1971, Harashima1972} which is used as a benchmark. 
However, vector-perturbation precoding is known to outperform Tomlinson--Harashima precoding, see, e.g., \cite{Windpassinger2004} for a comparison. 
Moreover, different block lengths are used in \cite{Pathapati2020} for the AE implementation and the benchmark precoder. 
We also note that the AE design in \cite{Pathapati2020} is different from ours in the sense that CSI is not provided as an input to the transmitter NN. 
Instead, the AE is trained and evaluated for the same fixed channel realization. 
As stated in \cite{Pathapati2020}, this has the downside that the AE needs to be retrained if the channel changes. 

\section{Interference Channel}
\label{sec:ic}

The last scenario we consider is the Gaussian interference channel, where $N$ transmitter--receiver pairs, each having a single antenna, communicate over the same physical channel. 

\subsection{Background and Baseline Schemes}

The interference channel is modeled by
\begin{align}
    \label{eq:ic}
    \vect{Y}_k = \mat{H}_k \vect{X}_k + \vect{N}_k, 
\end{align}
where $\mat{H}_k \in \mathbb{C}^{N \times N}$ is the channel matrix, $\mat{X}_k = [\vect{x}_{k,1}, \ldots, \vect{x}_{k,N}]^\transpose $, 
$\mat{Y}_k = [\vect{y}_{k,1}, \ldots, \vect{y}_{k,N}]^\transpose $,
and $\vect{x}_{k,i},\vect{y}_{k,i} \in \mathbb{C}^{N_B}$ are, respectively, the transmitted and received symbol vectors of the $i$-th user. 
As before, $\mathrm{vec}(\boldsymbol{N}_{k}) \sim \mathcal{CN}(\boldsymbol{0},N_0\boldsymbol{I}_{N N_B})$ is i.i.d.~Gaussian noise. 

As noted in \cite{OShea2017}, the optimal signaling scheme for the interference channel is a long-standing research problem. 
Existing approaches include, for example, superposition coding with private and common codebooks \cite{Han1981} or interference alignment \cite{Cadambe2008}. 
In this paper, we restrict ourselves to the same scenario as considered in \cite{OShea2017}, where $[\mat{H}_k]_{i,j} = 1$ for all $i,j \in \{1, \cdots, N\}$. 
Moreover, it is assumed that all users have the same average power constraint $\mathbb{E}\{\|\vect{x}_{k,i}\|^2\} = N_B P_T$ for $i = 1, \ldots, N$.  
A simple baseline scheme in this case is to use a time-sharing approach, where the transmitters send their messages in a round-robin fashion while all other transmitters remain silent.
This effectively orthogonalizes the interference channel into $N$ parallel and independent Gaussian channels. 
This baseline scheme was also considered in \cite{OShea2017} to benchmark the AE.

\begin{figure}[t]
    \centering
    \includegraphics[width=8.7cm]{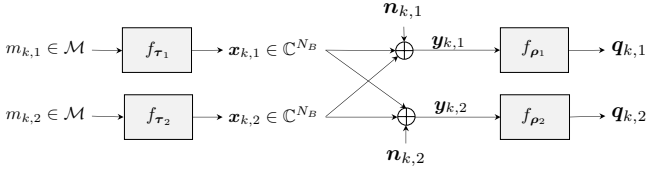}
    %\vspace{-0.1cm}
    \caption{Interference channel AEs, where two users communicate over the same physical channel. }
    \label{fig:icSYSTEM}
\end{figure}

\subsection{Autoencoder Design and Training}
\label{sec:ic_ae}

In the following, all users have the same message set $\mathcal{M}$. The generalization to different message sets for each user is straightforward. 
Each user maps their message $m_{k,i} \in \mathcal{M}$ to transmitted symbols via a transmitter NN $f_{\tau_i} : \mathcal{M} \to \mathbb{C}^{N_B}$  according to 
\begin{align}
    \vect{x}_{k,i} = f_{\tau_i}(m_{k,i}),
\end{align}
where we enforce $\mathbb{E}\{\|\vect{x}_{k,i}\|^2 \} = N_B P_T$ through a standard normalization layer, similar to \eqref{eq:normalization}. 
After all users have transmitted their symbols over the channel \eqref{eq:ic}, the receivers process the received symbol vectors $\vect{y}_{k,i}$ via an NN by generating $M$-dimensional probability vectors $\boldsymbol{q}_{k,i} = f_{\rho_i}(\vect{y}_{k,i})$ for $i = 1, \ldots, N$.
The loss function for user $i$ is the expected cross-entropy 
\begin{align}
    \label{eq:loss_ic}
    {\mathcal{J}}_i(\vect{\theta}) = 
    - \mathbb{E}\left\{  \log[f_{\rho_i}(\boldsymbol{y}_{k,i})]_{m_{k,i}} \right\}, 
\end{align}
where we use $\vect{\theta} = \{\tau_1, \ldots, \tau_N, \rho_1, \ldots, \rho_N\}$ to denote all transmitter and receiver NN parameters. 
Note that the expectation in \eqref{eq:loss_ic} is over the channel noise and the transmitted messages of \emph{all} users. 
\begin{figure}[t]
    \centering
    %[mode=build]
    % \includestandalone{tikz/interference_ch}
    \includegraphics[width=\columnwidth]{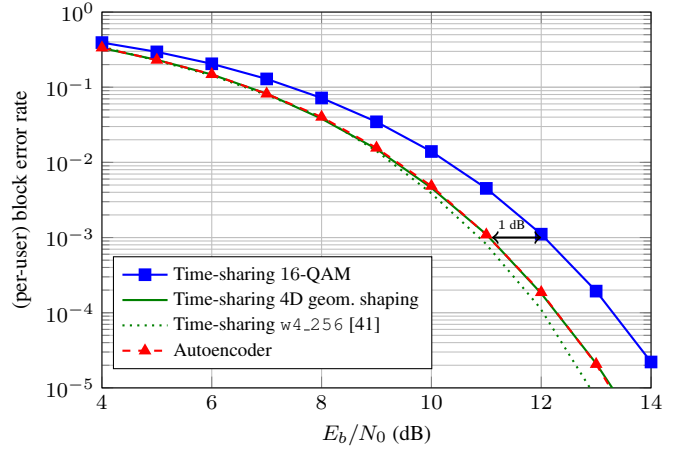}
    %\vspace{-0.1cm}
    \caption{BLER for the interference-channel AEs with $N = 2$, $N_B = 4$, $M=256$. The baseline corresponds to time sharing with three different modulation formats. } 
    \label{fig:ic}
\end{figure}

To optimize the parameters $\vect{\theta}$, a weighted average of the individual losses in \eqref{eq:loss_ic} for $i = 1, \ldots, N$ can be used. 
To encourage equal system performance among users, the weights can further be chosen dynamically in each gradient-descent iteration, where the weight for user $i$ is set proportionally to the corresponding per-user loss in the previous iteration \cite{OShea2017}. 
For example, the common loss function for $N = 2$ users in iteration $t$ is
%\begin{align}
%\label{eq:joint_ce}
$\mathcal{J}_{\text{CE}}
%(\vect{\theta}) 
= \alpha_t  \mathcal{J}_1 
%(\vect{\theta}) 
+ (1-\alpha_t) \mathcal{J}_2
%(\vect{\theta})
,$
%\end{align}
where 
\begin{align}
\alpha_{t+1} = \frac{
    \hat{\mathcal{J}}_1(\vect{\theta}_t)
}{
    \hat{\mathcal{J}}_1(\vect{\theta}_t) + 
    \hat{\mathcal{J}}_2(\vect{\theta}_t)
}, \qquad t > 0,
\end{align}
with $\alpha_0 = 0.5$ and we recall that $\hat{\mathcal{J}}_i$ refers to the Monte Carlo approximation of the expectation in \eqref{eq:loss_ic}. 
A block diagram of the AE setup for the interference channel is shown in Fig.~\ref{fig:icSYSTEM}.

\subsection{Numerical Results and Discussion}
\label{sec:ic_results}

We consider the case where $N=2$ users transmit over $N_B = 4$ complex channel uses and each user has a message set of cardinality $M = 256$. 
This corresponds to an uncoded transmission rate of $r = \log_2(M)/N_B = 8/4 = 2$ bits per channel use (bpcu) and user. 
The NN parameters are identical to the ones in \cite[Table~IV]{OShea2017} and also shown in Table~\ref{tab:network_parameters}.
Note that for this setup, no parameters are shared between any of the four NNs. 
Training is performed using the Adam optimizer with learning rate $0.001$ at $E_b/N_0 = P_T/(r N_0) = 11\,$dB (cf.~Table~\ref{tab:hyperparameters}). 
We use the normalized SNR $E_b/N_0$ for this scenario to make it easier to compare to prior work in \cite{OShea2017}.

Fig.~\ref{fig:ic} shows the performance of the trained AE (red triangles) in terms of the per-user BLER $\text{Pr}\{\hat{m}_{k,i} \neq m_{k,i} \}$ for the first user $i=1$, where the performance of the second user is essentially the same and omitted from the plot. 
As a comparison, the time-sharing baseline is shown, where the two users alternate $16$-QAM transmission (blue squares) which again gives a rate of $r=2$ bpcu and user.\footnote{These two cases are referred to as AE(4,8) and TS(4,8) in \cite[Fig.~6]{OShea2017}.}  
It can be seen that the AE outperforms this baseline by around $1\,$dB at a BLER of $10^{-3}$. 

\begin{figure}[t]
    \centering
    % \includestandalone{tikz/interference_rotation}
    \includegraphics[width=\columnwidth]{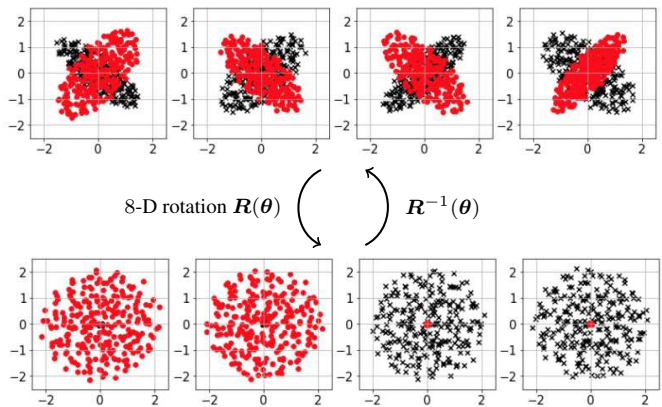}
    %\vspace{-0.1cm}
    \caption{Top: learned constellations for the interference channel ($256$ points per user), where different colors correspond to different users (cf.~\cite[Fig.~7 (d)]{OShea2017}). Bottom: the same constellations after applying an optimized rotation matrix (see the appendix for details). } 
    \label{fig:ic_rotation}
\end{figure}

The above results are consistent with the ones reported in \cite[Fig.~6]{OShea2017}. However, no explanation for the performance gain is provided in \cite{OShea2017}, where it is noted that the obtained results are ``difficult to interpret''. 
In the following, we aim to provide an explanation for the observed gains. 
First, we note that the baseline scheme can be improved by performing GS. 
In particular, since the two AEs jointly transmit messages over $N_B = 4$ channel uses, a time-sharing scheme with $2$ users may utilize $N_B/2 = 2$ complex channel uses, i.e., $4$ real dimensions.
In other words, rather than time-sharing $16$-QAM, a better baseline scheme is obtained by time-sharing an optimized $4$-dimensional modulation format. 
To that end, we trained a conventional AE for a standard AWGN channel with $M=256$ and $N_B = 2$, as explained in Sec.~\ref{sec:siso_ae}. 
The performance when using the resulting AE in a time-sharing fashion is shown in Fig.~\ref{fig:ic} by the solid green line. 
Interestingly, this baseline gives the same BLER as the AE for the interference channel. 

Indeed, we argue that this is not a coincidence and that the scheme learned by the two interference-channel AEs corresponds, in fact, to time sharing, albeit in a rotated reference frame. 
To see this, we plot the learned signal constellations for the two users in the top of Fig.~\ref{fig:ic_rotation} (which is similar to \cite[Fig.~7 (d)]{OShea2017}). 
As noted in \cite{OShea2017}, the learned constellation clouds resemble ellipses with orthogonal major axes and varying focal distances. 
We noticed that these elliptic shapes can be reproduced by applying a random $8$-dimensional rotation matrix to the time-sharing AE scheme. 
Moreover, it is possible to find a rotation matrix that de-rotates the learned constellations in the top of Fig.~\ref{fig:ic_rotation} such that essentially all signal energy for the two users is confined to orthogonal time slots. 
The resulting constellations are shown in the bottom of Fig.~\ref{fig:ic_rotation}. 
The details about how to obtain the underlying rotation matrix are given in the appendix.

Lastly, we note that optimized modulation formats in $4$ dimensions have been studied before (see, e.g., \cite{Welti1974}) and the baseline scheme can be further improved. 
The format with $M=256$ points in \cite{Welti1974} corresponds to the intersection of a $4$-dimensional lattice and a spherical bounding region. 
This constellation is also available in \cite{codes.se} denoted by {\tt w4\_256}. 
Its performance in a time-sharing scheme is shown by the dotted green line in Fig.~\ref{fig:ic}.
It can be seen that the lattice-based format outperforms all other schemes discussed so far, where the gain is quite significant at high SNR. 
At this point, it is important to stress that the cross-entropy minimization used for training the AE does not necessarily minimize the BLER. 
Instead, an AE trained with cross-entropy loss maximizes a lower bound on the mutual information (MI), see, e.g., \cite{Shen2018ecoc}. 
Indeed, it can be shown that the learned AE constellation for the time-sharing scheme achieves a higher MI than the lattice-based format {\tt w4\_256} over the standard AWGN channel \eqref{eq:awgn} at high SNRs. 
%Indeed, when using MI as a performance metric, the lattice-based format {\tt w4\_256} achieves $7.836$ bpcu for $E_b/N_0 = 11\,$dB over the standard AWGN channel \eqref{eq:awgn}, whereas the learned AE constellation for the time-sharing scheme achieves a (slightly) higher MI at $7.842$ bpcu. 

\section{Conclusions and Future Work}
\label{sec:conclusion}

In this work, we have evaluated several AE-based MIMO and MU communication systems in order to quantify and explain potential performance gains over fair benchmarks. 
The systems under consideration were open-loop MIMO, closed-loop MIMO, MIMO broadcast, and the interference channel. 
For all cases, the AE provides optimized mappings from messages to transmit vectors, as well as optimized detectors. 
For open-loop and closed-loop MIMO, we have shown that previously observed performance gains of the AE compared to the baselines can be partially attributed to geometric constellation shaping and optimized bit and power allocation. 
For MIMO broadcast, we have proposed a novel decentralized AE structure that performs close to nonlinear vector-perturbation precoding and significantly outperforms conventional ZF. 
Lastly, for the considered Gaussian interference channel, we have provided an interpretation of the learned AE-based communication scheme, thereby explaining the performance gains observed in prior work. 
In particular, we have shown that the AE learns a ``rotated'' time-sharing scheme. 

In general, our work has shown that, for a wide variety of different scenarios, AE-based communication systems have the potential of learning very good solutions without a priori knowledge about complex mathematical tools or communication-theoretic principles. 
On the other hand, our work has also highlighted the fact that such systems do not necessarily perform better than state-of-the-art benchmarks, provided that the benchmarks are properly chosen. 
A particular emphasis in this work was placed on selecting benchmarks that include known geometrically-shaped signal constellations, many of which are available in open databases such as \cite{codes.se}. 
Compared to previous work, the improved baseline schemes have allowed us to provide additional insights into AE-based systems and, in some cases, full interpretations of the learned communication schemes. 

For future work, we believe that there are several important aspects concerning the use of AEs which deserve further study:
\begin{itemize}
    \item \emph{Channel Models:} 
    Similar to related prior work, we have adopted \RevA{memoryless} channel models based on i.i.d.~Rayleigh fading and AWGN. 
    However, real wireless systems may follow a different fading model and suffer from additional impairments such as memory effects or nonlinearities caused by imperfect hardware. 
    For such systems, existing design approaches potentially operate far from optimality and AE-based methods may provide significant performance gains. 
    \RevA{However, the AE architecture and training method would need to be appropriately modified, e.g., using orthogonal frequency-division multiplexing (OFDM) in the case of memory effects.}
    
    \item \emph{Training Complexity:} 
    The considered AEs require a relatively large amount of training data, with large batch sizes, in order to converge to a good solution. 
    Improving the convergence speed would allow for the exploration of a larger parameter space, for example in terms of the NN architecture, potentially leading to performance improvements. 
    
    \item \emph{Implementation Complexity:} 
    Another important aspect is the computational complexity at runtime in practical implementations. 
    While a thorough evaluation of the implementation complexity (including the associated performance--complexity trade-off) is beyond the scope of this paper, we note that model-compression techniques such as NN pruning can be used to significantly reduce the number of computations (often without much loss in performance). 
    
    \item \emph{Scalability:} 
    With more transmit and receive antennas and/or more users, the complexity scaling of the corresponding NNs (e.g., in terms of layers) is currently unknown. 
    Moreover, the employed one-hot encoding scheme causes input and output sizes to grow exponentially with the number of antennas and rate. \RevA{This scalability issue may become even more severe when one considers  dispersive channels in combination with OFDM, leading to hundreds or thousands of parallel channels.%
    %To be more precise, one-hot encoding will be required for each subcarrier, and there are for example more than 3000 subcarriers in 5G.
    }
    Alternative embeddings \cite{Rodriguez2018} or multi-hot sparse categorical cross entropy could help alleviate the latter issue.  
    Both these issues affect training convergence (due to more trainable parameters) and runtime computational complexity. 
    
    \item \emph{Rate adaptation:} 
    The considered AEs have a fixed data rate, which limits possibilities for rate adaptation. 
    New NN architectures are needed to provide rate-adaptive transmission. 
    
\end{itemize}

\appendix
%\section{Rotation}

To de-rotate the learned signal constellations of each user for the interference channel in Sec.~\ref{sec:ic_results}, we start by constructing an overall $n \times n$ rotation matrix 
\begin{align}
   \mat{R}(\bm{\theta}) = \prod_{\substack{i,j \in [n] \\ i < j}} \mat{G}^{ij}(\theta_{ij}),  
\end{align}
where $n = 2 N_B$, $[n] \define \{1, 2, \ldots, n\}$, $\mat{G}^{ij}(\theta_{ij})$ is a Givens rotation matrix, and $\bm{\theta}$ is a vector of length $n(n-1)/2$ that contains all parameters, i.e., all individual rotation angles $\theta_{ij}$. 
Then, let $\mat{R}_\text{u}(\bm{\theta}), \mat{R}_\text{l}(\bm{\theta}) \in \mathbb{R}^{n/2 \times n}$ denote the upper and lower half of $\mat{R}(\bm{\theta})$ and define
\begin{align}
    \tilde{\mat{X}}_1(\vect{\theta}) &= \mat{R}_\text{l}(\vect{\theta}) \mat{X}_1, \\
    \tilde{\mat{X}}_2(\vect{\theta}) &= \mat{R}_\text{u}(\vect{\theta}) \mat{X}_2, 
\end{align}
where $\mat{X}_1, \mat{X}_2 \in \mathbb{R}^{n \times M}$ are the learned AE signal constellations, i.e., each column in $\mat{X}_1$ and $\mat{X}_2$ corresponds to one constellation point for the first and second user, respectively. 
Note that for the example in Sec.~\ref{sec:ic_results}, we have $n=8$ and $M=256$. 
Finally, $\bm{\theta}$ is optimized using conventional stochastic gradient descent with loss function
\begin{align}
    \label{eq:rotation_loss}
    \mathcal{J}(\vect{\theta}) = \|\tilde{\mat{X}}_1(\vect{\theta})\|^2 + \|\tilde{\mat{X}}_2(\vect{\theta})\|^2
\end{align}
and learning rate $0.001$. 
The individual angles of the Givens rotation matrices are randomly initialized assuming a uniform distribution over the interval $[0,2\pi]$. 
Note that the optimization outcome and the resulting rotation matrix are not unique because the constellation of each user can be arbitrarily rotated in $4$ dimensions without affecting the loss \eqref{eq:rotation_loss}. 

\balance

\bibliographystyle{IEEEtran}
\bibliography{references_fixed} 

\end{document}